\documentclass{article}
\usepackage{graphicx,float}
\usepackage{amssymb,amsfonts,amsmath,amsthm}
\usepackage{hyperref}
\newtheoremstyle{problemstyle}  
        { }                            
        { }                            
        {\sf}  
        {15pt}                         
        {\bfseries\itshape}            
        {\normalfont\bfseries:}        
        {.5em}                         
        {}            
\theoremstyle{problemstyle}
\newtheorem{problem}{Problem}[section] 

\hypersetup{%
linkcolor=black,%
citecolor=black,%
linkbordercolor={1 1 1},%
citebordercolor={1 1 1}%
}%

\usepackage{multicol}
\newcommand{\begtoc}{%
\vspace{6pt}
\hrule

\begin{multicols}{2}
{\small{\tableofcontents}}
\end{multicols}
}%
\begin{document}

\title{{\bf{{What is dimension?}}}}
\author{Somendra M Bhattacharjee$^{1,2}$}
\date{%
$^1$ Institute of Physics, Bhubaneswar 751005, India \\
$^2$ Homi Bhabha National Institute, Training School Complex, Anushakti Nagar, Mumbai 400085, India\\
{email:somen\string@iopb.res.in}
}%

\maketitle
\hrule
\vspace{6pt}

  This chapter explores the notion of ``dimension'' of a set.  Various
  {\it power laws} by which an Euclidean space can be characterized
  are used to define dimensions, which then explore different aspects
  of the set.  Also discussed are the generalization to multifractals,
  and discrete and continuous scale invariance with the emergence of
  complex dimensions.  The idea of renormalization group flow
  equations can be introduced in this framework, to show how the power
  laws determined by dimensional analysis (engineering dimensions) get
  modified by extra anomalous dimensions. As an example of the RG flow 
  equation,  the scaling of conductance by disorder in the  context of 
  localization is used. A few technicalities, including the connection 
  between entropy and fractal dimension, can be found in the appendices.

\vspace{6pt}

\begtoc
\hrule
\vspace{6pt}
 


\newcommand{\fighyp}{%
\begin{figure}[htbp]
\begin{center}
\includegraphics[scale=0.51,clip]{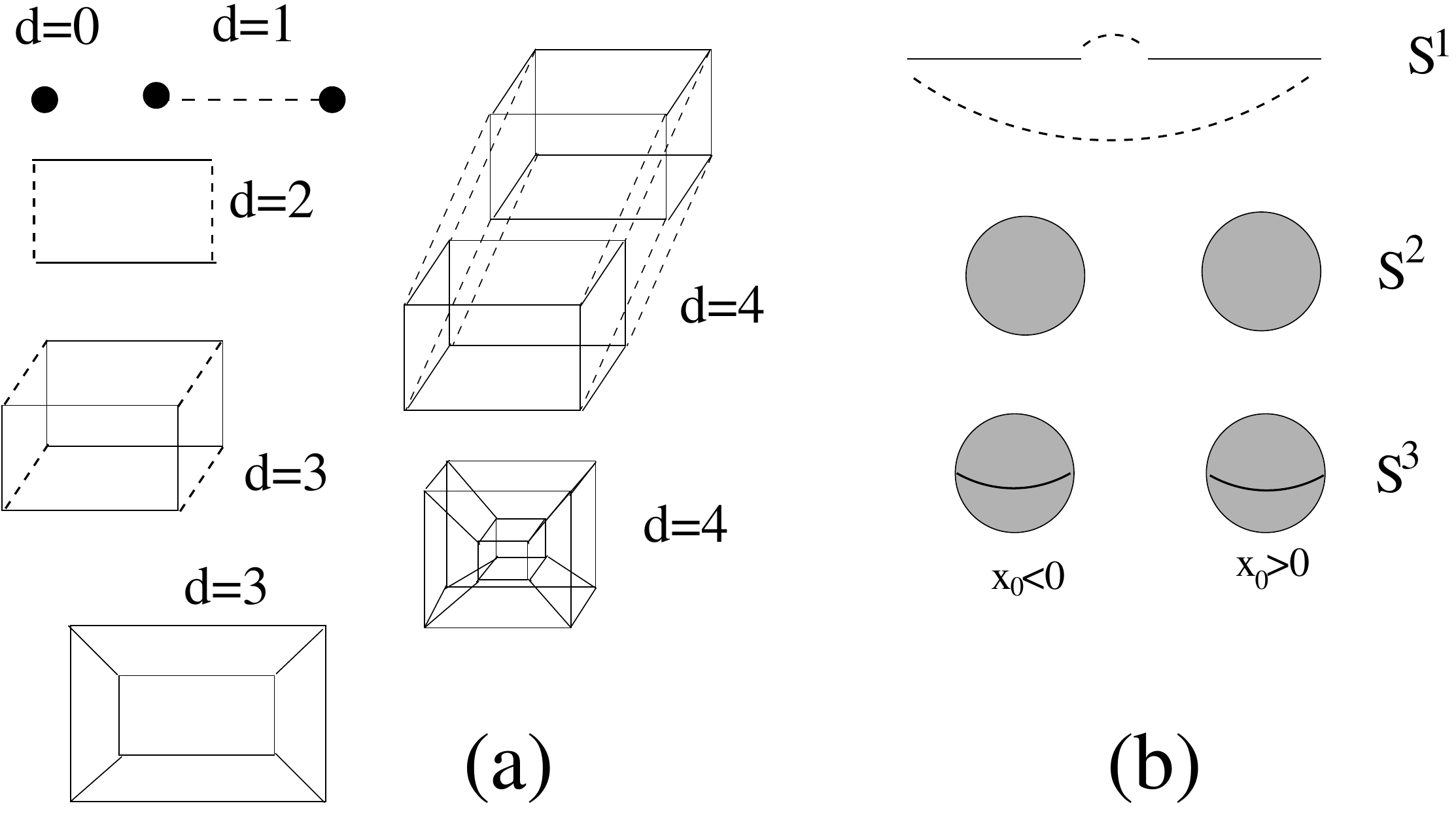}
\end{center}
\caption{(a) Construction of a $d$-dimensional cube (hypercube) by
  combining pairs of $(d-1)$-dimensional cubes.  The dotted lines are
  the new connections or bonds.  (b) Hyperspheres may also be
  constructed that way.  Take two closed intervals with the
  identification of the boundary points as indicated by the dashed
  lines to get $S^1$.  Similarly, two full spheres $B^d$, unit
  $d$-balls, one for $x_0<0$ and another one for $x_0>0$ with the
  identification of the boundaries gives $S^{d}$.  E.g., take two
  disks ($d=2$) and place one top of the other both centered at
  origin.  Since the boundaries are identified, puff the object (think
  of making {\it luchi} [https://en.wikipedia.org/wiki/Luchi]) giving
  an extra dimension as $x_0=\pm\sqrt{1-\sum_{i=1}^{d} x_i^2}$, the
  boundary being at $x_0=0$.  The net result is $S^{d}, $
  $\sum_{i=0}^{d} x_i^2 =1$. This construction is not possible for
  solid cubes or solid balls. }\label{fig:hype}
\end{figure}
}%

\newcommand{\figcant}{%
\begin{figure}[htbp]
\begin{center}
\includegraphics[scale=0.51,clip]{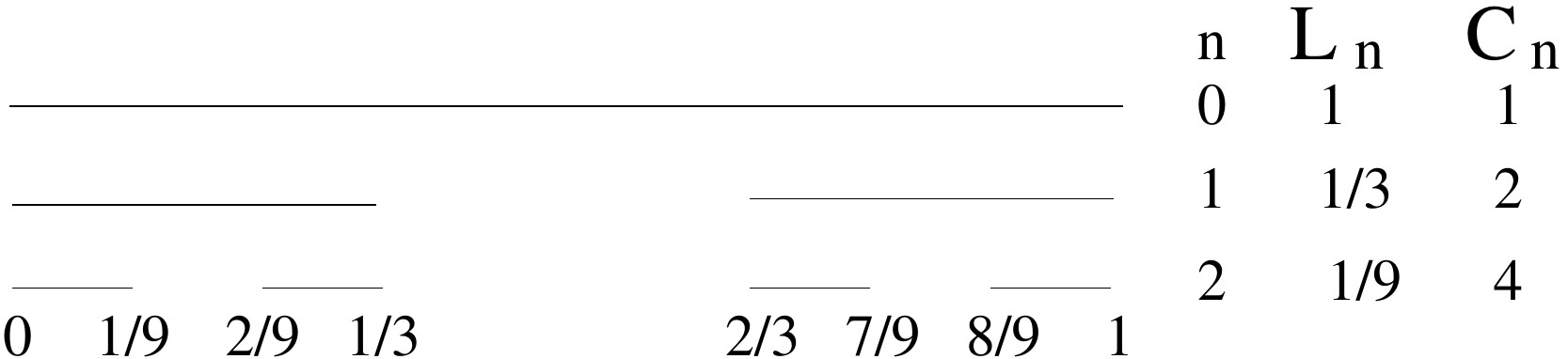}
\end{center}
\caption{Construction of a cantor set
}\label{fig:cant}
\end{figure}

}%



\newcommand{\myabst}{%
\begin{abstract}
\end{abstract}
}%


\section{Introduction}
The purpose of this chapter is to explore the idea of
``dimension'' of a set.  E.g., what does superscript 1 mean when
we talk of $S^1$?  The vector space idea of the minimum number of
basis vectors is too restrictive to be applicable to even many subsets
of standard Euclidean manifolds, which are not necessarily vector spaces.
Generalizations of the definition of dimension consistent with the
intuitions based on Euclidean spaces give a large number of
possibilities, with each one describing some aspect of the set.  These
dimensions need not be integers anymore.  If various physical
phenomena are studied on these subsets, which ``dimension'' will
matter?

Once released from the integer constraint, there is no restriction on
the value of dimension, with examples of positive, negative, real,
and complex, $d$, even cases which require more than one $d$.  Such
varieties are not possible in ordinary Euclidean spaces $\mathbb{R}^d$
because all definitions give the same $d$ for them.

It is amusing to note that the word ``dimension'' means several
different things in physics. It is used in statements like ``a line is
one dimensional'', and that is the ``dimension'' we are interested in
this chapter.\footnote{Common usages like ``love adds a new dimension
  to the conflict'' or ``the dimensions of this box are
  10cm$\times$20cm$\times$30cm'' can possibly be traced to the meaning
  of ``dimension'' as used in this chapter!}  The other one is the
dimension in the context of ``Units and Dimensions'', as for example,
force has dimension ${\sf M L T}^{-2}$, where ${\sf M, L, T}$ are the
dimensions of mass, length, and time respectively.  With the available
scales of a problem, like the interaction strength, thermal energy
($k_BT$), $\hbar$ (action), $c$ (velocity), we may, if we wish,
express all physical quantities in terms of length only.\footnote{As
  an example consider the path integral form of the propagator of a
  free particle, which involves a sum over all trajectories of
  $\exp(iS/\hbar)$, where $\hbar$ is the Planck constant divided by
  $2\pi$, $S= (m/2) \int dt\;  (d{\bf x}/dt)^2$ is the action, $m$
  and ${\bf x}$ are the mass and the position of the particle, $t$
  being time.  Since $S/\hbar$ is necessarily dimensionless, we may
  define $\tau=\hbar t/m$ to write $S/\hbar= (1/2) \int
  d\tau\; (d{\bf x}/d\tau)^2$.  With this form, $x$ has dimension
  ${\sf L}$ while redefined time $\tau$ has dimension ${\sf L}^2$.  }
The power of ${\sf L}$ we get by the dimensional analysis is to be
called the {\it engineering dimension} of the quantity.  One of the
aims of this chapter is to bring these two usages of ``dimension'' in
the same framework.  In the process we shall argue that the framework
allows ways to apparently ``violate'' dimensional analysis, and how
extra ``anomalous dimensions'' emerge.

The definitions of dimensions are based on various {\it power laws} by
which an Euclidean space can be characterized.  This procedure opens
up a new way of studying power laws, beyond geometrical objects, like
the divergences of response functions, e.g., susceptibility, near
critical points. The idea of renormalization group can be introduced
in this framework, to show, as just mentioned, how the power laws
determined by  engineering dimensions
get modified by extra anomalous dimensions.


  
\section{Does ``dimension'' matter?}
\label{sec:where-does-d}

Many phenomena when viewed in a broad way, are found
to depend on the dimensionality of the system.  Let's take a few
examples where $d$, the dimension of the space, occurs explicitly.
\begin{enumerate}
\item For noninteracting gases, classical or bosons or fermions, in
  $d$-dimensions, the thermodynamic fundamental relation is
  $PV=\frac{2}{d}\ U$ for pressure $P$, volume $V$, and total energy
  $U$.
\item Debye specific heat $c\sim T^d$ for a $d$-dimensional crystal at low temperatures $T$.
\item The probability$P_t$ of a random walker coming  back to the starting
  point in time $t$ satisfies $P_t \propto t^{-d/2}$. 
\item A diffusing particle has the characteristic mean square displacement
  $R^2\sim t$ in time $t$, in all {\it
    dimensions} so that the volume occupied is $t^{d/2}$. 
\item In quantum mechanics, a particle in a short range attractive potential may
  not have a bound state if $d > 2$, but there is always a bound state
  if $d<2$.
\item Take an Ising type model with short-range interactions.  It is
  known that there is a phase transition (critical point) if $d> 1$.
  For vector spins, a critical point exists only if $d>2$.  The
  critical behaviour is mean-field like if $d>4$.  Inbetween, the
  critical exponents depend on $d$ and a few other gross features.
\end{enumerate}
These are just a few.  What do we mean by $d$ in these statements?

\section{Euclidean and topological dimensions}
\label{sec:eucl-topol-dimens}

That a square lattice is two dimensional is easy to see if its vector
space property is known.  A cube, consisting of vertices and edges, can
be drawn on a piece of paper.
Fig \ref{fig:hype} shows possible constructions of
hypercubes and hyperspheres $S^n$ (drawn in $d=2$).  
\footnote{Convention: $S^n$ denotes the surface of an
  $(n+1)$-dimensional sphere while $B^n$ denotes an $n$-dimensional
  ball, i.e., a sphere with its interior and its boundary surface.
  For example $S^1$ is the set of points in two dimensions,
  $x^2+y^2=1$, while $B^2$ is the set of points with $x^2+y^2\leq 1$.
  This is equivalent to saying $S^1$ is the surface or boundary of
  $B^2$.}
That the cube is in some sense not a two  dimensional  object becomes
clear  if one wants to draw  on a plane larger lattices or graphs with such cubes
as units.

\fighyp

\subsection{Euclidean dimension}
\label{sec:euclidean-dimension}

A common procedure is to embed the lattice in an Euclidean space of
large enough dimensions.  From any point, draw a sphere of radius $R$
and count the number $N$ of points enclosed by the sphere.  Our
expectation is that, for a $d$-dimensional set, 
\footnote{A 
symbol $\sim$ indicates the functional relation without worrying about 
dimensional analysis, prefactors etc, while $\approx$ will be reserved 
for approximate equality.} 
$N\sim R^d$.
Exploiting this intuition, a definition of $d$ can be
\begin{equation}
  \label{eq:1}
  d=\lim_{R\to\infty}\frac{\ln N}{\ln R},\qquad ({\rm Euclidean})
\end{equation}
the asymptotic slope of a log-log plot of $N$ vs $R$. 
The above definition may be written  in a more useful form as
\begin{equation}
  \label{eq:16}
  R\; \frac{\partial N(R)}{\partial R}=d \; N(R),
\end{equation}
where   the coefficient of the linear term on the right hand
side corresponds to the dimension of the space.

An equation like Eq. (\ref{eq:16}) is suggestive of the form used in
renormalization group approach and can be linked to dimensional
analysis for various physical quantities.   If a physical quantity $A$, 
on dimensional grounds, depends on length as $L^c$, then there is an
equation equivalent to Eq. (\ref{eq:16}), viz.,
\begin{equation}
  \label{eq:18}
   L\;\frac{\partial A}{\partial L}=c\; A,
\end{equation}
where $c$ will be called the engineering dimension of $A$.

\subsubsection{$d$ via analytic continuation}
\label{sec:d-via-analytic}\index{dimension!analytic continuation}

In many problems, especially in renormalization group calculations
($\epsilon$-expansion), one generalizes the Euclidean dimension, Eq.
(\ref{eq:1}), to a continuous variable.  This is more of an analytic
continuation with the help of the metric than by any real construction
of any space.  A Gaussian integral in $d$ dimensions can be written as
\begin{equation}
  \label{eq:13}
  \int e^{-a r^2} d^d r=K_d\int_0^{\infty} e^{-a r^2} r^{d-1}  dr,
\quad{\rm with}\quad K_d=\frac{2\pi^{d/2}}{\Gamma(d/2)},
\end{equation}
as the surface area of a unit sphere.  An integration like the left hand 
side of Eq. \eqref{eq:13} above is
metric dependent.  Once converted to a one-dimensional  integral of a 
function where  $d$ appears as a parameter
(like the right hand side of Eq.\eqref{eq:13}), it is defined for any value of
$d$ allowing an analytic continuation to the whole complex plane of
$d$.  This is very useful to handle singularities or divergent
integrals (dimensional regularization) in many problems.  In this
analytic continuation approach, there is no association of any space
with noninteger $d$ and so we do not get into detailed discussion on
this approach in this chapter. (See Prob 3.1).

\subsection{Topological dimension}
\label{sec:topol-dimens}\index{dimension!topological}

It is possible to avoid any reference to an  embedding space by using
the intrinsic characteristics of the lattice.  By lattice we mean a
set of points connected by bonds and these bonds can be taken as a unit
or a scale for the connectivity of the points.  We take an $L$ step
path on the lattice from any one point and count all the new points
visited that were not seen upto the $(L-1)$th step.  This is like
counting the boundary points of an intrinsically defined sphere of
radius $L$.  Based on the expectation that  the boundary ``area''
grows like $N_b\sim L^{d-1}$,  the definition of the dimension is
\begin{equation}
  \label{eq:2}
   d_{\rm t}=1+ \lim_{L\to\infty}\frac{\ln N_b}{\ln L}. \qquad ({\rm topological})
\end{equation}
This will be called the {\it topological dimension} of the object.  It
is topological because this number does not change under continuous
deformation of the space.  In other words two homeomorphic spaces have
the same topological dimension.

A formal definition of the topological dimension is via the covers.  A
crude definition is that if $\delta$ is the dimension of the space
(or lowest dimension of all possible spaces) that separates 
our space into disconnected pieces, then the
topological dimension of our space is $1+\delta$.

We make a convention that a null set $\emptyset$ has dimension $-1$
while a point set has dimension $0$.  All others follow from the above
rule.

There are ambiguities.  As an example, take a line.  A line is broken
into two pieces by removing a point.  A point by definition is of zero
dimension.  Hence a line is a one-dimensional object.  This is however
a bit tricky.  If we think of a line in three dimensions, it can be
broken into two pieces by another line or by a plane etc.  In such
situations, we need to choose the smallest dimensional object to
determine $\delta$.  Another example could be a figure
$\Large{\overline{\bullet}}$ ( a line and a disk).
Being disconnected, a null set separates them, and, therefore, the
dimension should be $1+(-1)=0$.  But individually these are 1 (line) and 2
dimensional (disk) spaces.  In such a situation we define
a local dimension and choose the largest one.  In this particular case
it will be $d=2$.

The above rules may be formalized by an iterative procedure with the
basis sets of the space.  (a) If the boundaries of the basis sets are
of dimensions $\leq d-1$, then the space is of dimension $\leq d$.
(b) If this is true for $d$ but not for $d-1$, then the space has
dimension $d$.

For the disk-bar example above, the basis sets for the bar has
boundaries $d=0$ while the disk has basis sets with boundaries $d=1$,
i.e., the dimension of the boundaries of the basis sets satisfy $d\leq
1$.  Therefore $d\leq 2$.  Invoke (b) to rule out $d=1$ or any number
greater than 2.

Let us take $R$ with the usual topology defined by the open sets
$(a,b), b>a$.  The boundaries are 0-dimensional points for all such
open sets.  Hence $R$ has dimension $d=1$.  With inherited topology
for $S^1$, the basis sets (open arcs of a circle) have boundaries of
dimensionality $0$.  Therefore $S^1$ has $d=1$.

All the definitions used so far would give the dimension of $R^d$ to
be $d$.  This number $d$ happens to be the number of independent
vectors needed to span the space when viewed as a vector space.  The
dimensionality of a topological space is a topological invariant in
the sense that if there is a continuous mapping or homeomorphism that
takes $R^m$ to $R^n$, then $m=n$.

\begin{problem}
  A problem on dimensional regularization.  Show that the
  one-dimensional integral $I_1(x)=\int_{-\infty}^{\infty} \frac{d
    z}{\sqrt{x^2+z^2}}$ is  divergent.

To tackle this divergence, generalize the  integral to $d$ dimensions as
\begin{equation}
  \label{eq:14}
  I_d(x)=\int_{-\infty}^{\infty} ...\int_{-\infty}^{\infty} 
  \frac{d^d r}{\mu^{d-1}\sqrt{x^2+r^2}}=K_d
    \int_0^{\infty} \frac{r^{d-1} d r}{\mu^{d-1}\sqrt{x^2+r^2}},
\end{equation}
where an arbitrary length $\mu$ is introduced to maintain the correct
dimensions (engineering dimension!).  Formally, $I_d$ is $I_1$ for $d=1$.  The form on the
right hand side can be defined for any $d$.  With $d$ as a continuous
variable,  the integral is divergent\footnote{Important here is the
  behaviour at the upper limit, which we may see by putting a cutoff
  $L$ as $I_d^{(L)}\sim
  \int^L r^{d-2} dr\sim L^{d-1}/(d-1),$ for $d\neq 1$.  For $d=1, I_{d=1}^{(L)}\sim \ln
  L$.  Therefore $I_{d\geq1}\to\infty$ as $L\to\infty$.  In such
  analytic continuations of integrals,  log divergences are always very special.} for $d\geq
1$ but convergent for $d<1$.  This signals the possibility of a
singularity in the complex $d$-plane at $d=1$.   By doing the integral
in the convergent domain in the $d$-plane,  show, by using Gamma
functions and analytic continuation,  that
\begin{equation}
  \label{eq:15}
  I_d(x)= \left(\sqrt{\pi} \frac{x}{\mu}\right)^{d-1}\
  {\Gamma\left(\frac{1-d}{2}\right)}
 =\frac{2}{\epsilon}-2 \ \ln(x/\tilde{\mu}) +O(\epsilon),
\end{equation}
where $\epsilon=1-d$ is a small parameter for the expansion.  Absorb
some O(1) factor in $\mu$ to define $\tilde{\mu}$.

This particular example appears in the calculation of the electrostatic
potential due to an infinitely long uniformly charged wire.
\end{problem}

\begin{problem} (Mathematical) 
  Show that the definition of $d$ obtained by using a basis is independent of the
  basis chosen.

  One way of doing it is to define $d$ without any basis but with the
  help of the open sets.  The iterative definition of the dimension of
  a topological space $T$ for a set $X$ would be as follows: The
  dimension is $d$ if for any point $x\in X$ and any set $U\in T$
  containing $x$, there exists an open $V$ with $x\in V$ such that the
  closure of $V$ is contained in $U$ with dim$(\partial V)\leq d-1$.
  If this is satisfied for $d$ but not for $d-1$, then the dimension
  of the space is $d$.  It is assumed that dim$(T)$=-1 if
  $X=\emptyset$.
\end{problem}

\begin{problem}(Mathematical)
  Prove that if $R^m$ is topologically equivalent (i.e. homeomorphic)
  to $R^n$, then $m=n$.  This means the dimension is a topological
  invariant.
\end{problem}

\section{Fractal dimension: Hausdorff, Minkowski (box) dimensions }
\label{sec:hausd-dimens-fract}

Let us now look at some nontrivial examples.  We shall see that
topological dimension ($d_{\rm t}$)  is not enough; a few others are needed.  First
among these is the Hausdorff dimension ($d_{\rm f}$) that one gets by embedding the
set in a real space $R^n$ of appropriate dimension  $n$.

In the following the Cantor set (defined below) is taken as a
paradigmatic example because of its apparent simplicity.  It is an
example of a space of topological dimension 0 but it is not just a
finite collection of points.

\subsection{Cantor set: $d_{\rm t}=0, d_{\rm f}<1$}
\label{sec:cantor-set}
There are many ways to define Cantor sets.  The middle $1/3$ rule used
below is historically the first one defined by Cantor and will be
called {\it the} Cantor set.

\figcant

The set is constructed iteratively by taking a closed interval 0 to 1,
and then removing the middle 1/3 to get two {\it closed } intervals.  In the
next step, the middle one third of the two branches are removed
leaving us with 4 intervals (Fig. \ref{fig:cant}). This iterative
process leaves a set of disconnected points,
\begin{eqnarray}
  \label{eq:4}
\left. \begin{array}{lcl}
S_0&=&\left[0,1\right],\\
S_1&=&\left[0,\frac{1}{3}\right]\cup\left[\frac{2}{3},1\right],\\
S_2&=&\left[0,\frac{1}{9}\right]\cup\left[\frac{2}{9},\frac{3}{9}\right]\cup\left[\frac{6}{9},\frac{7}{9}\right]\cup\left[\frac{8}{9},1\right],\\
...
\end{array}\right\}\quad {\cal S}=\cap_{n=0}^{\infty} S_n.
\end{eqnarray}
 
First note that the lengths of intervals removed are successively
$\frac{1}{3}, \frac{2}{3^2},...$ so that the total length removed is
$\sum_n \frac{2^{n-1}}{3^n}=1$.  Since the total length we started
with is $1$, the remaining points have no ``length'' and therefore
cannot be one-dimensional.  The set is disconnected.  Therefore the
{\it topological dimension} is $d_{\rm t}=0$.  Despite this
0-dimensionality, there is still a nontrivial structure like
self-similarity.  If any one part of the $n$th iterate is multiplied
by $3$, we get back the state in the $(n-1)$th iterate.  This
geometric structure is described by a different dimension, to be
called the self-similarity dimension or box dimension or Hausdorff
dimension.

For a regular object like the Cantor set, the pattern is obtained by
scaling the $n$th iterate by a scale $b<1$ and combining $C_b$ of
them.  Here, $b=1/3$ and $C_b=2$. On successive rescaling how the
number grows is given by the Hausdorff dimension\index{dimension!Hausdorff}
\begin{equation}
  \label{eq:20}
  d_{\rm H}=\frac{\ln (C_{b})}{\ln (1/b)}.\qquad {\rm (Hausdorff )}
\end{equation}

For the Cantor set, $d_{\rm H}=\frac{\ln 2}{\ln 3} \approx 0.63 <1$.  We
have assumed a power law dependence, $C_b\sim b^{-d_{\rm H}}$.

Another practical  procedure is to cover the set by boxes.    Consider
the set as a part of $R^d$.  Cover it by
boxes of length $L$ and count the number of boxes occupied by the set.
Let this be $B_L$.  If the length is changed to $bL$ by a scale factor
$b<1$, the number changes to $B_{bL}$. The {\it box dimension} is
then defined as\index{dimension!box}\index{dimension!Minkowski}
\begin{equation}
  \label{eq:3}
  d_{\rm f}=\lim_{n\to\infty} \frac{\ln (B_{b^nL}/B_L)}{\ln (1/b^n)}.\qquad
  {\rm (Box\;\; or\  Minkowski )}
\end{equation}
By using successive generations, we may also write the above equation
as
\begin{equation}
  \label{eq:57}
  d_{\rm f}=\lim_{n\to\infty} \frac{\ln (B_{b^nL})
    -\ln(B_{b^{n-1}L})}{\ln (1/b^n) - \ln(1/b^{n-1})},\qquad
  {\rm (Box\;\; or\  Minkowski )}
\end{equation}
which is the discrete version of Eq. (\ref{eq:16}).

For the Cantor set, if we choose $b=1/3$, then  $B_{b^n}=2^n$, and,
therefore, $d_{\rm f}=\frac{\ln 2}{\ln 3} =\log_3 2=d_{\rm H}$.  The box dimension is also
called the {\it Minkowski dimension}.

The occurrence of $\ln 3$, or log base $3$, is not accident. It comes
from the scale $1/3$ under which the Cantor set is scale invariant.
If we choose an arbitrary scale factor, say $b=1/2.9$, the scale
invariance of the Cantor set is not obvious.  This existence of a
special scale, here $1/3$ or its powers, is an example of {\it
  discrete scale invariance}.  In Appendix we discuss how a discrete
scale invariance leads to complex dimensions.\index{discrete scale invariance}

\begin{figure}[h]
  \begin{minipage}[c]{0.25\textwidth}
    \includegraphics[width=\textwidth]{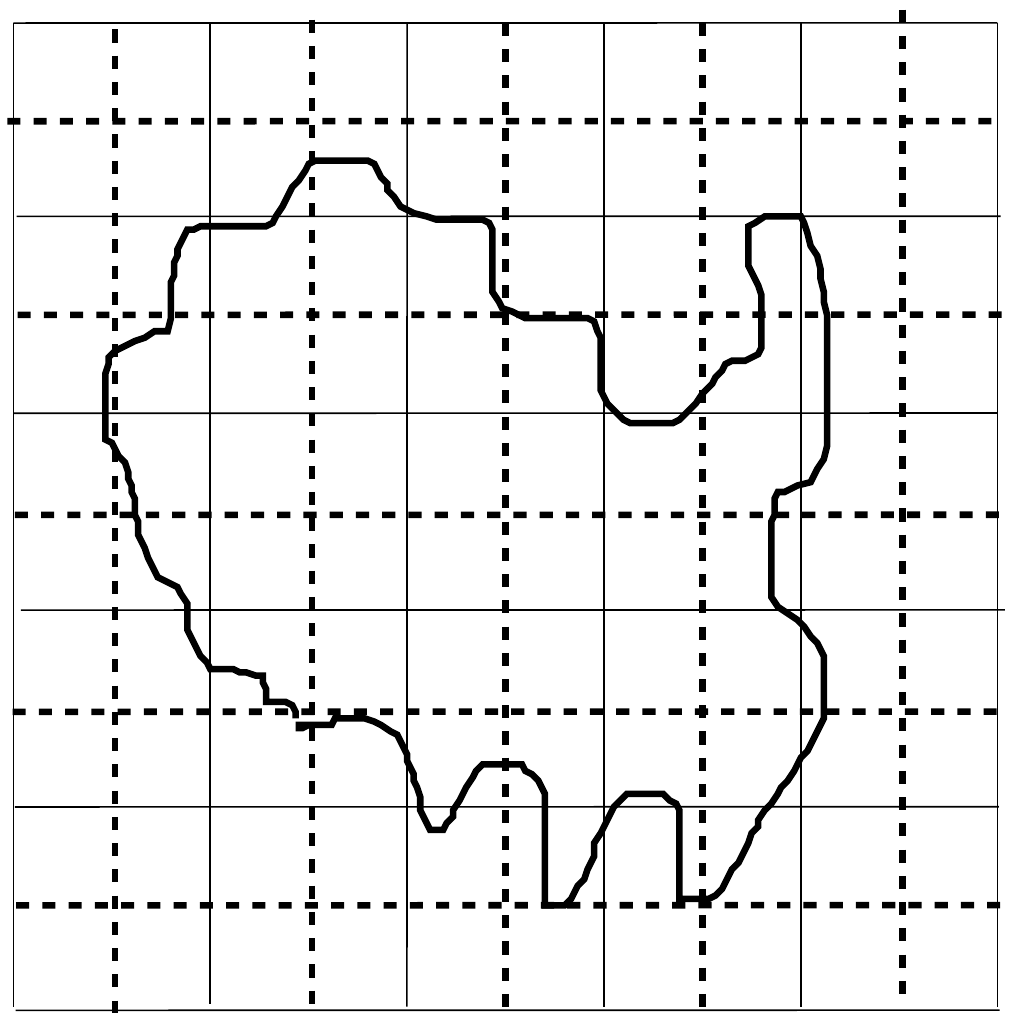}
  \end{minipage}\hfill
\quad  \begin{minipage}[c]{0.7\textwidth}
  \caption{Covering by boxes.  Count the number of boxes required to
    cover the object for different box sizes like those represented by
    the solid grid and the dotted grid.  The dotted grid is finer and
    therefore would involve larger number of boxes.  }
    \label{fig:box}
  \end{minipage}
\end{figure}

Whenever the Hausdorff dimension is different from
the topological dimension, the set is called  a
{\it fractal}.\index{fractal}
  For most regular fractals, $d_{\rm H}=d_{\rm f}$, but there are cases
where they may differ.  When they are same, they may be called the
fractal dimension or the scaling dimension.

A subtle difference in the way the Hausdorff and the Minkowski
dimensions are defined may be noted here.   The Hausdorff dimension is
obtained by going to the large size limit by scaling up the structure,
while the Minkowski dimension is from the opposite limit.  In the latter
case, one explores to the short scale behaviour by using progressively smaller
boxes.  From an experimentalist's point of view, the Hausdorff
dimension is obtained when probed in the long wavelength limit, called
the infrared limit, while the Minkowski (box) dimension is obtained
when probed using shorter  wavelengths, called the ultraviolet limit.
 
\subsubsection{When Hausdorff$\neq$ Minkowski}
\label{sec:when-hausd-mink}

An example of a case of different Hausdorff and Minkowski dimension is
the set of rational numbers in $(0,1)$.  If we want to cover this set
by linear ``boxes'', every box will contain some points.  It follows
from the fact that the rationals form a dense set in $(0,1)$. Therefore
the box dimension or Minkowski dimension is one.  On the other hand,
each rational number is an isolated point with dimension zero.  A {\it
  countable union } of points will then have a Hausdorff dimension of
zero.  Generally if $F$ is a dense subset of an open region of
${\mathbb R}^n$, then its box dimension is $n$.  Note however that
Cantor set does not belong to this category because it is {\it
  uncountable}.

\subsubsection{Inhomogeneous scaling}
 \label{sec:inhom-scal}
\index{inhomogeneous scaling}
 
Let us write the definition of $d_{\rm H}$ in Eq. (\ref{eq:20}) in a
different way as $C \lambda^{d_{\rm f}}=1$, where $\lambda (=1/b)<1$ is the
scale factor and $C$ is the number of such scaled objects combined
to generate the next generation.  This equation may now be extended
to a situation of inhomogeneous scaling where each of the $C$ objects
has its own scale factor $\lambda_i<1$.  The fractal dimension is then
the solution of the equation $\sum_{i=1}^{C} \lambda_i^{d_{\rm f}}=1$.

\subsubsection{Fractal dimension as a Continuous variable}
\label{sec:fractal-dimension-as}
\index{dimension!continuous}

Instead of the 1/3 rule of the Cantor set, we may remove any open
interval  $(x,1-x),
x<1/2$.  After that we remove the appropriate $(1-2x){\cal L}$ part from each
of the remainders of length ${\cal L}$.  Let's call it th $x$-rule.
  For $b=x$,  $C_{b^n}=2^n$.
The fractal dimension is 
\begin{equation}
  \label{eq:17}
 d_{\rm f}(x)=\frac{\ln 2}{\ln (1/x)},\qquad {\rm(continuous)} 
\end{equation}
with $d_{\rm f}(x=0)=0, d_{\rm f}(x=1/2)=1$ as
expected.  It is therefore possible to construct a family of fractals
with fractal dimension $d_{\rm f}$ as a continuous variable in the range
$[0,1]$.

Fractals of dimension $d_{\rm f}>1$ can be constructed by taking advantage
of product spaces.  Given a $1<d_{\rm f}<m$, where $m$ is an integer ($>1$), 
use the $x$-rule  to construct a fractal ${\cal F}$ of
$d_{\rm f}^{\prime}=d_{\rm f}/m<1$.  Then construct the direct product space
${\cal F}\times ...\times{\cal F}$.  For the scale factor $b$, we now
need $m$-dimensional boxes so that the number of boxes covered is
$C_{bL}^m$.  The fractal dimension, by Eq. (\ref{eq:3}), is 
$m \; d_{\rm f}^{\prime}=d_{\rm f}$.

Since $m$ can be any integer greater than the chosen value of $d_{\rm f}$, we
end up with many different product spaces all of the same fractal
dimension $d_{\rm f}$.  Therefore, $d_{\rm f}$ is a necessary but not a sufficient
characterization of the space.

\subsubsection{Configuration space of the Ising model}
\label{sec:conf-space-ising}
\index{Ising model}
We raised the question of the configuration space of the Ising model
which consists of $N$ spins $s_i=\pm 1$.  For each spin we have a
discrete topological space $\{0,1\}$ so that for infinitely many spins
arranged in a one-dimensional lattice, the total configuration space
is a product space $\{0,1\}^Z$. 
We now show that the configuration space can be mapped on to a set of
real numbers $\in[0,1]$, whose fractal dimension can be determined.

Let us first consider a particular case using  ternary expansion of numbers.  By
construction, any member of the Cantor set can be expressed as
\begin{equation}
  \label{eq:60}
y=\sum_{n=1} \frac{a_n}{3^n},\quad{\mathrm{where}}\quad a_n=0\; {\rm or} \; 2.  
\end{equation}
This is because every
point in $S_n$ exists in all the previous generations.  
Therefore  any  point can be tracked  as belonging to either  the 0th or  the 2nd
interval of the previous step.\footnote{ Warning: there are two
  possible representations of some numbers like $\frac{1}{3}=\sum_n
  \frac{2}{3^n}$, i.e., in base 3 notation, $(10000...)$ and
  $(2222...)$ denote the same number. We omitted the decimal point in front
of the numbers.  Similarly
  $\frac{1}{4}=\frac{2}{9}+\frac{2}{9^2}+...=\frac{1}{3}-\frac{1}{3^2}+\frac{1}{3^3}-\frac{1}{3^4}+...$.
  In such cases we choose the representation involving 0 and 2. This
  restriction to $0$ and $2$ makes the binary string to Cantor set a 
 one-to-one and onto mapping.}
Equivalently, the points can be represented by an infinite string
$(a_1a_2a_3....a_n....)$, like $(02220220002....)$ , or by dividing by
2, an infinite string of 0's and 1's.  For  an infinitely long chain
of Ising spins, a configuration $(s_1s_2...s_n...)$ can be converted
to a real number
\begin{equation}
  \label{eq:61}
  x=\sum_{n=1}^{\infty}\; \frac{2 s_n}{3^n}, \quad {\rm with}\quad s_n=0,1,
\end{equation}
similar to Eq. (\ref{eq:60}).
 In other words, the topological
space of the Cantor set can be mapped on to $\{0,1\}^Z$, or,
equivalently,  can be mapped onto the
configuration space of an  infinitely
long Ising chain.  

The entropy per spin of the Ising model in the high temperature limit
is $k_B \ln 2$ where $k_B$ is the Boltzmann constant
What we learn from this analysis is that the entropy (in this case the
high temperature entropy) of the Ising system (or, for that matter, any two
state model) is determined by the dimension $\log_3 2$ of the
set of real numbers equivalent to the configuration space .  The
connection between the dimension of the equivalent set of real
numbers and the entropy is discussed in Appendix A where we also
show that  base $3$ is nothing special.

\begin{problem}
The Ising two state problem can be mapped on to  the Cantor set as
the collection of all infinite strings of 0 and 1 occurring with equal
probability.
Suppose, instead, 0 occurs with probability $p$, and 1 with $q=1-p$. The
physical entropy per spin is known to be $s=-k_{B} (p \ln p +q \ln q)$, where
$k_B$ is the Boltzmann constant.
Show that the fractal dimension of the set of real numbers  is $d_{\rm f} \propto S$,
because $d_{\rm f}=-(p \ln p +q \ln q)/\ln 3.$ 
\end{problem}
\begin{problem}
A generalization of the above problem is to consider the set of all
strings of base $m$ numbers, i.e.,  strings of $0,1,...m-1$.  A
string $\{a_n\}, a_n=0,1,...m-1$, corresponds to a real number
$x=\sum_{1}^{\infty} a_n/m^n$.  If the digits $0,1...m-1$ occur with
probabilities $p_n, n=0,...,m-1$, then the fractal dimension of the
et consisting of $x$'s is $d_{\rm f}=- (\sum_n p_n \ln p_n )/\ln m$.

A consequence of this result is that if only $s_n=0,1$ occur with
probabilities $p,q=1-p$,  then $d_{\rm f}= -(p\ln p+q\ln q)/\ln m$,,
e.g., for the set $x=\sum_{n} (m-1)s_n/m^n$ (a generalization of the
standard Cantor set).

\end{problem}

\begin{problem}
  Consider the infinitely long Ising chain configurations, but now
  with a restriction that no two 0's can be adjacent (or nearest
  neighbours).  This occurs in nonabelian anyon chains discussed in later
  chapters. (See Ref. \cite{preskill}).  For $N$ spins, the total
  number of configurations is not $2^N$ any more.  If $C_N$ is the
  number of $N$ spin configurations under this
  restriction, then show that $C_N=C_{N-1}+C_{N-2}$.\\
  Hint: If the first spin is 1, then the second onwards can be any of
  the allowed $N-1$ spin configurations.  This is $C_{N-1}$.  If the
  first one is 0, then by restriction, the next one has to be 1 but
  the spins are free after that.  The number of such configurations is
  $C_{N-2}$.  Note that $C_1=1, C_2=3$.  This is the Fibonacci sequence.

  If $C_N\sim \tau^N$ for large $N$, then show $\tau=(1+\sqrt{5})/2$
  (golden mean), as expected for the Fibonacci numbers.  Show that the
  fractal dimension of the corresponding real number set is $\propto
  \ln \tau$ (see problems 4.1 and 4.2).
 
A number like $\tau$ here, different from the
  standard value $2$, is often called the {\it quantum
    dimension}\index{dimension!quantum} of the anyonic chain,
  \index{dimension!quantum}.  Suppose we consider spin-1/2 particles.
  For each spin the Hilbert space is 2 dimensional.  Then the
  Hilbert space for $N$ spins is the  tensor product with dimensions $2\times
  2\times 2...=2^N$.   In contrast for the (Fibonacci-) anyon chains, even though
  individually the spaces are two dimensional, the  dimension
  of the $N$-anyon Hilbert space is $\tau^N$ for large $N$.  This is
  as if the effective 
  dimension of individual Hilbert space is $\tau$.  To  recognize this
  difference, this dimension is called ``quantum dimension''.   It is
  interesting to note that the topological entanglement entropy is
  determined by $\ln \tau$.   This goes beyond the scope of this chapter. 
\end{problem}

\begin{problem}
  What is the configuration space of Ising spins on a square lattice?
  Explore if there is any mapping to real numbers, or any connection
  with the fractal dimension of the set of numbers, as found for a one
  dimensional chain of spins.
\end{problem}

\subsection{Koch curve: $d_{\rm t}=1, d_{\rm f}>1$}\label{sec:koch-curve}

\begin{figure}
\includegraphics[width=\textwidth,clip]{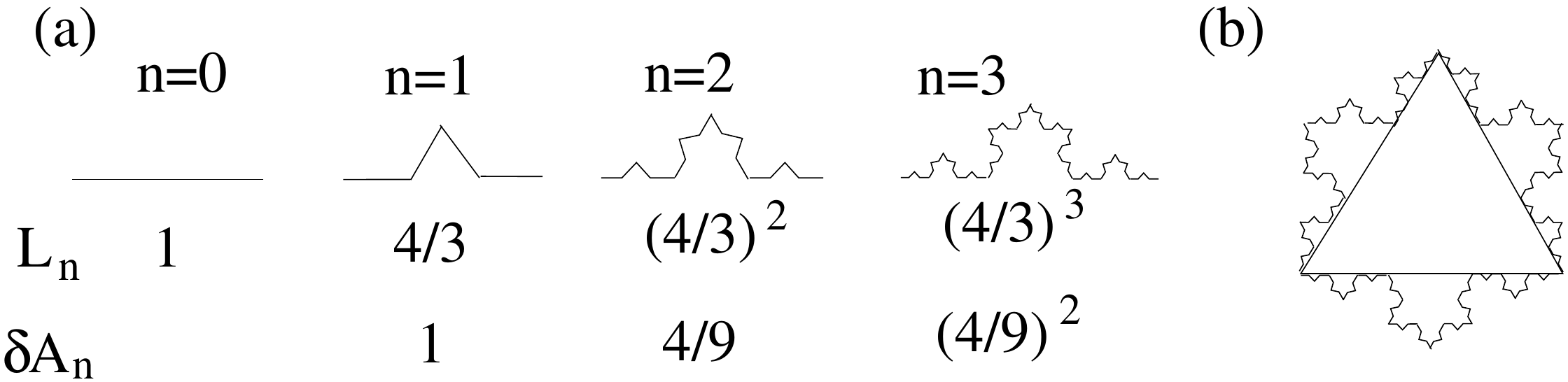}
\caption{(a)Iterative construction of the Koch curve. An interval of
  length 1 is divided into 3 equal parts.  The middle segment is
  replaced by an equilateral triangle of side length 1/3 without the
  base.  This procedure is repeated for each segment.  The length of
  the curve ($L_n$) and the area ($A_n$) under it for the first few
  generations are given.  (b) Similar construction with a triangle.
  After infinite iteration we get a closed loop of infinite perimeter
  but of finite area.  This is equivalent to $S^1$ where ``1'' refers to
  the topological dimension. }\label{fig:koch}
\end{figure}

A Koch curve is defined in Fig. \ref{fig:koch}. \index{Koch curve}
\index{curve!Koch} Instead of deleting the middle $1/3$ as in the
Cantor set, we add an extra piece increasing the length of the line.
This curve has the following properties:
\begin{enumerate}
\item A point disconnects it.  Therefore it has a {\it topological dimension}
 $d_{\rm t}= 1$.
\item The generation-wise lengths are $1, \frac{4}{3},\left(
    \frac{4}{3}\right)^2,...\left( \frac{4}{3}\right)^n,...$ so that
  the length $L_n\to\infty$ as $n\to\infty$.  However the area under
  the curve is $1+ \frac{4}{9}+\left(
    \frac{4}{9}\right)^2+...=\frac{9}{5}$.
\item For a stick length $l_n=1/3^n$, the number of sticks is
  $C_n=4^n$. For a scale factor $b=1/3$, the ratio of the two numbers
  $C_{n+1}/C_n=4$.  The
  fractal dimension is $d_{\rm f}=\lim_{n\to\infty} \frac{\ln
    (C_n/C_{n-1})}{\ln (1/b)}=\frac{\ln 4}{\ln 3}$
\end{enumerate}
A practical procedure is to cover the curve with a square grid of unit
length $l=1$ and count the number of boxes $C_l$ occupied by the
curve.  Then change the grid size by a scale factor $b$ and count
$C_{bl}$. One may then use the slope of the log-log plot with  the
definition of Eq. (\ref{eq:3})
to determine $d_{\rm f}$.

If we take $\epsilon=1/3^n$ as the length of the measuring stick with
$\epsilon\to 0$ as $n\to\infty$, then the measured length $(4/3)^n$ is
dependent on the scale via $n= -\ln\epsilon/\ln 3$.  We may generalize
this result.  The length measured at scale $\epsilon$ behaves as
  \begin{equation}
    \label{eq:5}
L(\epsilon)\stackrel{\epsilon\to0}{=}L_0 \epsilon^{-\alpha}, \quad
{\rm with\ }  \alpha=d_{\rm f}-d_{\rm t}.
  \end{equation}
For those cases where the two dimensions match (as in $R$), the length
is independent of the scale. In such cases, one may talk of  the length
of the curve, and such curves are called {\it rectifiable
  curve}.\index{rectifiable curve}\index{curve!rectifiable} 

We see a curve of topological dimension 1 but of a fractal dimension
between 1 and 2.  Koch curve is also an example of a continuous but nowhere
differentiable curve. As a closed curve,  Fig. \ref{fig:koch}b, we get
a continuous curve enclosing a finite area, though of infinite
length.   Fig. \ref{fig:koch}b is therefore topologically equivalent
to $S^1$, where we now recognize the superscript as the topological
dimension of the boundary.

\begin{problem}
  Construct a space filling curve, i,e, a curve of topological
  dimension 1 but of fractal dimension 2.  An example is the  Peano
  curve. See Ref. \cite{listwiki}.\index{Peano curve}\index{curve!Peano}
\end{problem} 
\subsection{Sierpinski Gasket: $d_{\rm t}=1, d_{\rm f}>1$}\index{Sierpinski
  Gasket}\index{fractal!Sierpinski gasket}
\label{sec:sierpinski-gasket}
\begin{figure}
\includegraphics[width=\textwidth,clip]{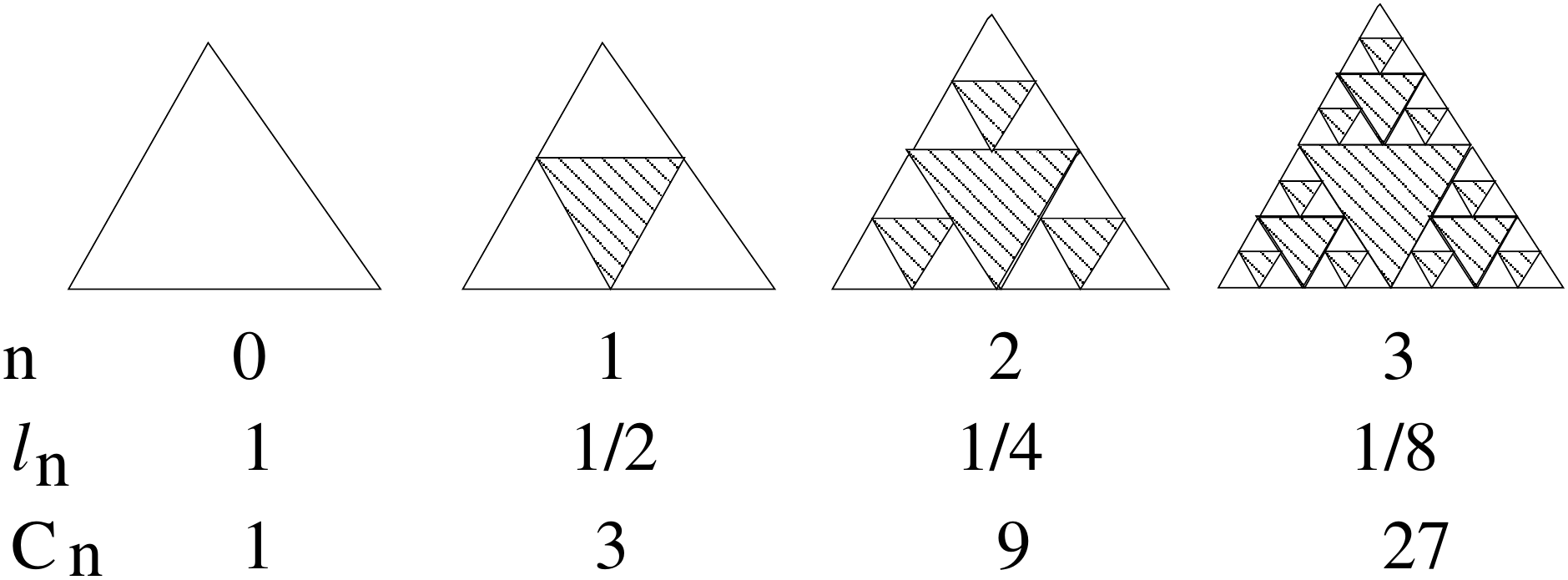}
\caption{(a)Iterative construction of the Sierpinski gasket.  One way
  is to view it as an aggregation of triangles so that the side length
  increases by a factor of 2, with the inner one missing (shaded
  triangles).  Another view is to punch holes and remove the inner 1/3
  of every triangle.  The length for counting ($l_n$) and the number
  ($C_n$) of triangles are noted underneath.}\label{fig:sierp}
\end{figure}

The construction of the Sierpinski gasket is shown in Fig.
\ref{fig:sierp}.  This fractal can be disconnected by isolated points
and is therefore topologically one-dimensional. 

A fractal that can be disconnected by a finite set of points is called
a finitely ramified fractal.   A finitely ramified fractal is therefore
has a topological dimension 1.     The fact that we need 3 copies at a
scale factor of 2 tells us that the fractal dimension of the Sierpinski gasket
is $d_{\rm f}=\frac{\ln 3}{\ln 2}$.

An intuitive way of arguing that its fractal dimension is less than 2
is to note that there 
are holes at every scale, and therefore holes will be present no
matter at what resolution we look at, unlike a compact object.  This of course requires
$n\to\infty$.  

\begin{problem}
Show that the  Sierpinski gasket is topologically equivalent to $S^1$
but it is not rectifiable (i.e., its length $\to\infty$).\index{curve!rectifiable}
\end{problem}

\begin{problem}
Sierpinski Carpet: Take a square of side length 1. Divide each
side\index{fractal!Sierpinski Carpet}
into three pieces of length 1/3 and remove the inner square of area
1/9.  Repeat this process ad infinitum.

(a) Show that the total area is zero.

(b)  Show that the carpet consists of points with a ternary expansion
$(\sum_{n=1}^{\infty} \frac{a_n}{3^n},\sum_{n=1}^{\infty}
\frac{b_n}{3^n})$ where $(a_n,b_n)\in {\cal S}$, ${\cal S}=
\{0,1,2\}\times \{0,1,2\}-\{(1,1)\}$.  The topological space is ${\cal
  S}^{Z_+}$. 

(c)  Identify  the  Cantor sets along the diagonal and the
medians.\index{Cantor set}

(c) Show that the fractal dimension is $d_{\rm f}=\frac{\ln 8}{\ln 3}$.

(d) Show that it is an infinitely ramified fractal, but with topological
dimension = 1. Note that the topological dimension by definition is an
integer.  For the carpet, it has to be less than $2$, and it is not
$0$.  Hence it is $1$.  Construct a direct proof of this.

(e) Show that the Sierpinski Gasket can be homeomorphically embedded in
 the Sierpinski carpet.  Prove  a more general statement: ``any Jordan
 curve\footnote{A Jordan curve is a planar simple closed curve
   homeomorphic to a circle.  Simple here means nonintersecting.}  can
 be homeomorphically embedded in the Sierpinski carpet.''  This was
 proved by Sierpinski in 1916.

\end{problem}

\begin{problem}
{Hierarchical lattices}\index{Hierarchical lattice}
\begin{minipage}[c]{0.47\textwidth}
\begin{center}
\includegraphics[scale=0.75,clip]{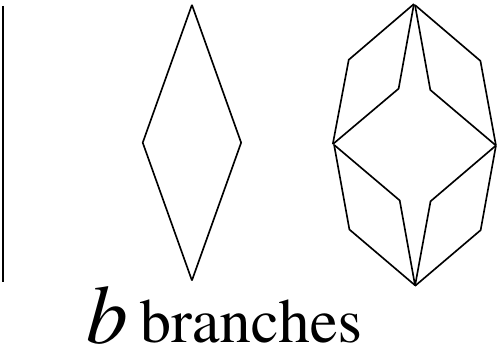}
\end{center}
\end{minipage}\hfill

  A hierarchical lattice is built by successive replacement of a bond by a
  motif.  In the example the motif consists of a diamond like object
  with $b$ branches and $2b$ bonds.  This is also called a diamond
  hierarchical lattice.  Show that the fractal dimension is
  $d_{\rm f}=\frac{\ln 2b}{\ln 2}$.

The hierarchical construction of such lattices helps in easy
implementation of renormalization group transformations or scaling.
\end{problem}

\subsection{Paths in Quantum mechanics:$d_{\rm t}=1, d_{\rm f}=2$}\index{Quantum mechanics!paths}
\label{sec:quantum-mechanics}
We now argue that the trajectories of a nonrelativistic
quantum\index{quantum mechanics}
particle is a fractal obeying Eq. (\ref{eq:5}).  The traditional
Brownian motion also belongs to the same class.  

From the  uncertainty principle, fluctuations in position ($\delta x$) and
momentum ($\delta p$) are related by   
$\delta x\;\delta p\sim \hbar$ and $\delta p= m \delta x/\delta t$,
($t$ being time), it follows that
\begin{equation}
  \label{eq:7}
(\delta x)^2\sim \delta t.  
\end{equation}
Take these $\delta x, \delta t$  as  the scales for space and
time to measure the length of a trajectory from $(x_1,t_1)$
to $(x_2,t_2)$.  
The time interval $T=t_2-t_1$  consists of  $N=T/\delta t$ pieces so that
the length is 
\begin{equation}
  \label{eq:6}
 L=N\ \delta x=\frac{T}{\delta t} \delta x= \frac{T}{\delta x},
\end{equation}
where Eq. (\ref{eq:7}) has been used.  Therefore, $L\to\infty$ as
$\delta t\to 0$.  We conclude, by comparing with Eq.  (\ref{eq:5}),
that the fractal dimension of the trajectory is $d_{\rm f}=2$, though
a path, by definition, has a topological dimension $d_{\rm t}=1$.

\begin{problem}
  For the cases where the propagator in the path integral approach can
  be calculated exactly, it is observed that only the classical path
  contributes.  By definition, a {\it classical path} has $d_{\rm f}=d_{\rm t}=1$.  Show
  that this happens because of the interference of the nearby {\it quantum
  paths}.
\end{problem}

\section{Dimensions related to physical problems}
\label{sec:dimens-relat-phys}
To go beyond topology and geometry, we need to study some physical
problem on a fractal.  These could be of several types as mentioned in
Sec. \ref{sec:where-does-d}.  Let's consider those cases one by one.

\subsection{Spectral dimension}\index{dimension!spectral}
\label{sec:spectral-dimension}

A physical way to explore a space is to use a probe that would in
principle involve the whole space.   In Euclidean space, a few such
probes are (1) diffusion processes or random walks, (2) elastic waves
or lattice vibration and (3) any quantum mechanics problem.  The
common link among  the three types is via the Laplacian  in the
Euclidean space as follows.
\begin{enumerate}
\item   The diffusion process is described by the differential
  equation for any diffusing field $\psi$,\index{diffusion}
\begin{equation}
  \label{eq:24}
  \frac{\partial \psi}{\partial t}=D \nabla^2 \psi,
\end{equation}
where $D$ is the diffusion constant.
\item  The free particle  Schr\"odinger equation is described
  by\index{Schr\"odinger equation} 
\begin{equation}
  \label{eq:25}
  i \hbar  \frac{\partial \psi}{\partial t}=- \frac{\hbar^2}{2m} \nabla^2 \psi,
\end{equation}
where $m$ is the mass of the particle and $\hbar$ is the Planck constant
divided by $2\pi$.
\item  The wave equation, e.g., describing sound waves is\index{wave equation}
\begin{equation}
  \label{eq:26}
   \frac{\partial^2 \psi}{\partial t^2}=  c^2\nabla^2 \psi.
\end{equation}
\end{enumerate}
The diffusion equation also occurs in heat transport and defining the
Laplacian for any space is often called the heat-kernel problem.  For
some of the fractals, it is easier to consider the lattice vibration
problem or the scalar version the resistor problem in the zero
frequency limit.   Th dimensionality of the space can then be defined
by the appropriate generalization of the Euclidean results.  This we
do below.  The dimension we obtain in this way is called the {\it
  spectral dimension} $d_s$.

\subsubsection{Diffusion, random walk}
\label{sec:diff-rand-walk}
For a diffusion problem in $\mathbb{R}^d$\footnote{ For a random walk on a
lattice, the Gaussian distribution of Eq. (\ref{eq:8}) is valid for
lengths.} the probability distribution of the end to end vector ${\bf
  R}$ is a Gaussian\index{random walk}
\begin{equation}
  \label{eq:8}
P({\bf R})=\frac{1}{(2\pi t)^{d/2}}\ \exp\left(-\frac{d
  R^2}{2t}\right)\equiv  \frac{1}{(2\pi t)^{d/2}}\ f(R/\sqrt{t}).
\end{equation}
In this limit short scale details are not
expected to be important.
A general form allowing a dimensionality dependence is to write it as
\begin{equation}
  \label{eq:9}
P({\bf R}) \sim  { t}^{-d_s/2}\ f(R/\sqrt{t^{1/d_w}}),
\end{equation}
defining a walk dimension $d_w$ and a spectral dimension $d_s$.
For $\mathbb{R}^n$, $d_s=n, d_w=2$.
Since probability is normalized,  where the integration over ${\bf R}$
involves the fractal dimension, a change of variable gives 
$$\int d^{d_{\rm f}}R \; P({\bf R})\sim t^{d_{\rm f}/d_w - d_s/2}.$$  Since the
power of $t$ should be zero, it follows that 
\begin{equation}
  \label{eq:10}
  d_s=2\frac{d_{\rm f}}{d_w}.
\end{equation}
Only for $d_w=2$, the  Hausdorff and the spectral dimensions
match.
This turns out to be the case for hierarchical lattices of Prob. 3.6.
 
Based on Eq. (\ref{eq:9}), the spectral dimension can be defined from
the return to origin probability $P(0,t)$ as
\begin{equation}
  \label{eq:27}
  d_s=-2 \lim_{t\to\infty} \ \frac{\ln P(0,t)}{\ln t}.
\end{equation}
As for the paths in quantum mechanics (Sec.
\ref{sec:quantum-mechanics}), $d_w$ is the fractal dimension of the
walk.  It may also be seen as the scaling relation between space
rescaling and time rescaling as determined by the dynamics.  In such
contexts, $d_w$, denoted by $z$,  is called the dynamic exponent.  For
diffusion like processes  $z=2$.  The dynamic
exponent is an important characteristic quantity near various  phase
transitions.

\subsubsection{Density of states}
\label{sec:density-states}

For a quantum mechanical problem with energy dispersion relation
$\omega =C k^{p}$ for small wave vector $k=|{\bf k}|$ in $d$
dimensions, the density of states for $\omega \to 0$ is $\rho(\omega)\sim
\omega^{-1+d/p}$, with the integrated density of states as $I(\omega)\sim
\omega^{-d/p}$.  For the electronic problem with a quadratic
dispersion relation, $p=2$ and one gets $\rho(\omega)\sim \omega^{-1+d/2},
I(\omega)\sim \omega^{d/2}$.   For phonons, $p=1$ and so, $\rho_{\rm
  phonon}(\omega) \sim \omega^{d-1}$, which gives the Debye law for specific heat.
The low energy excitations involve $k\to 0$ (long wavelength) for
which small scale details do not matter.  In other words, the large
scale dynamical features of the fractal or the object are probed by the
long wavelength propagation.
As a generalization, $d$ in the above densities may be replaced
by $d_s$ which is called the spectral dimension.

It is shown in Appendix B that $d_s$ for the Sierpinski Gasket is
different from its fractal dimension by explicitly calculating the
spectral dimension.

\section{ Which $d$?}
\label{sec:answer-what-d}

We now go back to the questions asked at the beginning of this chapter,
Sec. \ref{sec:where-does-d}.

\subsection{Thermodynamic equation of state}
\label{sec:them-equat-state}
The thermodynamic relation alluded to at the beginning actually comes
from the density of states.   In the grand canonical ensemble,  the
grand potential (the equivalent of free energy) is $-PV$, and for a
noninteracting gas, it can be
written as
\begin{equation}
  \label{eq:51}
  PV= k_B T \int d\epsilon \; \rho(\epsilon) \; \ln {\cal{Z}}(\epsilon,\beta,\mu),
\end{equation}
where $\epsilon$ is the energy, $\rho(\epsilon)$ the density of
states, $\beta=1/k_BT$ the inverse
temperature, $\mu$ the chemical potential, and ${\cal{Z}}$ is the
single state grand canonical partition function, 
\begin{equation}
  \label{eq:53}
  {\cal{Z}}=\sum_{n=0}^{\infty}  g_n \exp[-\beta(\epsilon-\mu)n],  
\end{equation}
where
  \begin{equation}
    \label{eq:54}
g_n=\left\{\begin{array}{cl}
                                     1,& {\mathrm{for\  bosons,}}\\
                                     (n\string!)^{-1},& {\mathrm{for
                                         \ classical \ particles,}}\\[4pt]
                                     \left[\begin{array}{ll}
                                             1,& {\mathrm{if}}\  n=0,1\\
                                             0,& {\mathrm{if}}\  n>1
                                             \end{array}\right.  & 
                                           {\mathrm{for\ fermions,}}
                                           \end{array}
                                           \right.
\end{equation}
though the explicit form is not required for our argument.

The average particle number  and the average energy are  then given by
\begin{equation}
  \label{eq:52}
  \langle N\rangle= \int d\epsilon\; \rho(\epsilon)\;
  {\overline{n}}(\epsilon,\beta,\mu), 
\quad {\mathrm{and}} 
\quad U= \int    d\epsilon\; \rho(\epsilon)\; \epsilon \;{\overline{n}}(\epsilon,\beta,\mu),
\end{equation}
  where
\begin{equation}
  \label{eq:55}
{\overline{n}}(\epsilon,\beta,\mu)=k_BT \frac{\partial\ln    {\cal{Z}}}{\partial \epsilon}.
\end{equation}
We now take the definition of spectral dimension to write
$\rho(\epsilon)=A\epsilon^{-1+d_s/2}$ for free particles of mass $m$
with a dispersion relation $\epsilon=p^2/2m$, with $A$ some constant.
The density of states may be divergent but is always integrable.  With
these information in hand, an integration by parts of the integral for
$U$ in Eq. (\ref{eq:52}), in conjunction with the (negative) grand
potential in Eq. (\ref{eq:51}) gives us the required relation
\begin{equation}
  \label{eq:11}
  PV=\frac{2}{d_s} U.
\end{equation}
The thermodynamic fundamental relation for an ideal gas involves the
spectral dimension of the space.

\subsection{Phase transitions}
\label{sec:phase-transitions}
The notion of lower critical dimension arose from studies of symmetry
breaking in various dimensions.  See Ref. \cite{smbcrit} for an
introduction to critical phenomena. 

Two different cases are to be
considered, namely a discrete or a continuous symmetry breaking.  To
be concrete, it is better to consider a particular case.  A typical
Hamiltonian is $H=-J\sum_{<ij>} {\bf S}_i\cdot {\bf S}_j$ where ${\bf
  S}_i$ is a unit $n$-component spin vector at site $i$ of say a
hypercubic lattice.  For the Ising model $S_i=\pm 1$ is just a
discrete variable with $n=1$. Continuous cases correspond to $n\geq
2$; the planar xy model has $n=2$, the three dimensional spin $n=3$ is
the Heisenberg ferromagnet, etc.  The Hamiltonian allows an ordered
ground state where all spins are parallel.   Therefore at zero
temperature ($T=0$) we get  a ferromagnetic state that breaks the
rotational invariance of $H$.   Note that $H$ remains invariant under
a rotation of the spins by any amount if performed on all the spins
(global symmetry).   For the Ising case, the symmetry is discrete with
$S_i\to - S_i$.   As we raise temperature, thermal fluctuations tend
to destroy the perfect alignment of spins due to entropic reasons. 
Therefore the question arises whether the broken symmetry state
persists at nonzero temperatures.   It is known that at very high
temperatures entropy wins yielding a paramagnetic phase.   In case the
ordered state persists, then there has to be a special temperature
$T=T_c$ (called the Curie point) at which a phase transition takes
place.   If we look at this particular $H$ in various dimensions, then
$T_c\neq 0$ only for $d>d_l$, where $d_l$ is called the lower critical
dimension.
For the Ising case, $d_l=1$ while for $n\geq 2$, $d_l=2$.  For a
fractal, to which $d$ are we referring?

\subsubsection{Discrete case: Ising model}
\label{sec:discrete-case:-ising}
The Landau-Peierls argument for the Ising Hamiltonian is a generic way
of determining the lower critical dimension for discrete symmetry
breaking.  Let us start with an ordered state, say all up spins at
$T=0$, and isolate a domain of opposite spins (down spins).  By
symmetry, the two states have the same energy, and, therefore, the cost of
flipping the spins is in the creation of the boundary.  Furthermore,
same number of down spins can be enclosed by boundaries of different
shapes.  The boundary therefore has an entropy associated with it.
Since the boundaries define the topological dimension of the system,
we may write the change in free energy as
\begin{equation}
  \label{eq:63}
\Delta F= \sigma L^{d_{\rm t}-1}-Ts_0 L^{d_{\rm t}-1},   
\end{equation}
where $\sigma$ is the energy cost of creating a unit ``area'' boundary
and $s_0$ is the associated entropy.  E.g., for the Ising case on a
square lattice, $\sigma=2J$.  The free energy expression in Eq.
(\ref{eq:63}) is valid for $d_{\rm t}>1$ for which such flipped
domains may not destroy the ordered state if $T<\sigma/s_0$.  Thus a
ferromagnetic state with broken symmetry can exist at nonzero
temperatures.  If $d_{\rm t}=1$, the cost in energy is independent of
the size but the flipped block can be placed anywhere on the lattice
giving an entropy $\propto\ln N$ where $N$ is the size of the system.
For large $N$, entropy dominates, and so the system goes over to a
paramagnetic state at any nonzero $T$.  {\it The lower critical
  dimension is therefore one, and it is the topological dimension that
  matters}.  This means the Ising model on a Sierpinski Gasket does
not show a ferromagnetic state.

The above simple picture is not complete, because we have seen that
the topological and the fractal dimensions are not sufficient to
characterize all fractals.  For example, the Ising model does not show
any phase transition for the Sierpinski Gasket, a finitely ramified
fractal, but does show a transition on the Sierpinski carpet, an
infinitely ramified fractal, even though both fractals have
topological dimension 1.  With many parameters, the lower critical
dimension loses its significance.  In any case, one may safely say
that, for discrete symmetry, the condition of topological dimension
$\leq 1$ is necessary but not sufficient for {\it no} symmetry
breaking transition.\index{symmetry breaking}

\subsubsection{Continuous case: crystal, xy, Heisenberg }
\label{sec:continuous-case:-xy}

Let us first consider the case of crystal, and ask the question
whether the crystalline state, a state of continuous symmetry
breaking, can survive in presence of lattice vibrations due to thermal
fluctuations.  Let ${\bf u}({\bf r})$ be the small displacement of the
particle at site ${\bf r}$ of the crystal.  The thermally averaged
correlation function $C(r)=\langle {\bf u}(0)\cdot {\bf u}({\bf
  r})\rangle$ tells us if the crystal state can be defined.  In case
this correlation does not decay to zero for large separation $r$, the
long range order of the state gets destroyed.  The independent
vibration modes are the Fourier modes, ${\bf u}({\bf k})$, with a
dispersion relation $\omega_k = v |k|$, at least for small $k$, in
Euclidean spaces.  As independent oscillators, by equipartition
theorem, $\langle u(k)^2\rangle\sim k_BT/\omega^2$ with $\omega$ in
the range $(0,\omega_{\rm max})$.  The real space correlation then
involves an integral over all the modes,
$$ C(r)\sim \int \rho(\omega) \; \frac{1}{\omega^2}\; d\omega.$$
With $\rho(\omega)\sim \omega^{d_s-1} $,  the correlation seems to
diverge for $d_s\leq 2$.   The divergence comes from the low frequency part
which corresponds to vibrations spanning large distances, thus
exploring the real space.
It is therefore  the {\it spectral dimension} that matters.  The symmetry
is restored by thermal fluctuations if $d_s\leq 2$.  Note
that $d_s=2$ is excluded for the ordered state to exist.

Similar arguments can be used for the xy or the Heisenberg magnets.  
It is still
important to know why the arguments for the discrete symmetry cannot 
be applied here.
As the order parameter space is continuous and  
connected\footnote{See S. M. Bhattacharjee, {\it Use of Topology in physical problems},  arXiv:1606.04070.}, 
there is no well defined
domain wall separating the states.  Even if we start with a sharp wall,
the variations near the wall can be smoothened-out to make a thicker wall.
The thicker the wall, the less costly it is, invalidating the
Landau-Peierls argument for discrete symmetry.

\subsection{Bound states in quantum mechanics}
\label{sec:bound-states-quantum}

Let us consider a particle in an attractive short range potential
well.  We know from explicit solutions that though a bound state is
guaranteed in low dimensions, it is not so in higher dimensions.  What
is the borderline dimension?\index{quantum mechanics}

In a path integral approach, the trajectories spending 
a large fraction of time in the well contribute to the propagator for
a bound state.  These paths also involve excursions in the classically
forbidden region.  But once it is out of the well, it must come back
for a bound state.  The overall distance spanned (generally measured
by the root mean squared distance) in the classically forbidden region
determines the width, $\xi$, of the bound state wavefunction.  The
larger the width, the smaller is the bound state energy.  By
uncertainty principle, the energy, $E$, is given by $|E|\sim
1/\xi^2$, with $E\to 0$ as $\xi\to\infty$.  The question therefore is
tantamount to asking whether $\xi$ can be infinite even when the well
is not vanishing.  

Suppose, we want the propagator or the Green function for the particle
from the center of the well (origin) to origin, $K(0T|00)$ in time
$T\to\infty$.  Let $K_b(0t_2|0t_1)$ and $K_c(0t_2|0t_1)$ be the
propagators for paths inside the well and in the classically forbidden
region from time $t_1$ to $t_2$, and $v$ be the tunneling coefficient.
Then, treating time as a discrete variable (for simplicity)
\begin{equation}
  \label{eq:58}
  K(0T|00)=K_b(0T|00)+  \sum_{t_1,t_2} K_b(0T|0t_2)v K_c(0t_2|0t_1)vK_b(0t_1|00)+... .
\end{equation}
Generating functions can be introduced as
$$G(z)=\sum_t K(0t|00)z^t,G_b(z)=\sum_t K_b(0t|00)z^t,G_c(z)=\sum_t K_c(0t|00)z^t,  $$
so that Eq. (\ref{eq:58}) for $T\to\infty$  can be written as a geometric series
\begin{eqnarray}
  \label{eq:59}
  G(z)&=&G_b(z)+G_b(z) v^2 G_c(z) G_b(z)\nonumber\\
&&  +G_b(z) v^2 G_c(z) G_b(z) v^2 G_c(z) G_b(z)+...\nonumber\\
&=&\frac{G_b(z)}{1-v^2 G_c(z) G_b(z)}.
\end{eqnarray}
The singularity coming from the denominator of Eq. (\ref{eq:59})
determines the quantum bound state energy, while the singularities of $G_b$
and $G_c$ determine the energies of the classical bound state and the
unbound state.
 
The important quantity here is then the return probability, that a
particle going from the origin comes back to origin in time $t$.  The
probability of returning to origin in time $t$ is given by $K_c\sim
t^{- d_s/2}$, as we saw in Sec. \ref{sec:diff-rand-walk}.  $G_c(z)$ is
then determined by $d_s$.  A direct analysis of the singularities of
$G(z)$ shows that a bound state with excursions in the classically
forbidden region may not exist if $d_s>2$.  We refer to Ref.
\cite{smsmb} for details.  In short, the existence of a bound state is
determined by the spectral dimension of the space.

\section{Beyond geometry: engineering and anomalous dimensions}
\label{sec:scale-invar-power}

It is possible to go beyond geometric figures, and use the ideas of
the previous sections in a more broader context.  Any function $f(x)$
is to be called scale invariant if $f(x)=b^{\eta}\ f(bx)$, under a scale
transformation $x\to bx$.  If this is true for any $b$, we may choose
$b=1/x$ to get $f(x)=x^{-\eta} f(1)$, a pure power law.  Most often
scale invariance and power laws are used synonymously.\footnote{In contrast to
the examples discussed earlier which had a discrete scale invariance
(only particular values of $b$ are allowed), this is a case of continuous
scale invariance.  This distinction is important. }

\subsection{Engineering dimension}
\label{sec:engin-dimens}

In order to distinguish a pure power law from other types, let us
consider a few special cases like,
\begin{equation}
  \label{eq:28}
f_2(x)=\frac{1}{(x+a)^{c}},\quad f_3(x)=e^{-x/a},\quad {\rm and\;}\
f_4(x)=\frac{e^{-x/a}}{x^{c}}.\quad (x>0)
\end{equation}
None of these functions show scale invariance in the true sense, but  can
be written as
\begin{equation}
  \label{eq:30}
 f_j(x)=x^{-c}\; {F}_j\left(\frac{x}{a}\right), \quad (j=2,3,4),
\end{equation}
with $c=0$ for $f_3$, and $F_j$ another function.  Such forms are called
scaling forms and can be arrived at by a dimensional analysis.  Taking
$x,a$ as lengths, if $f_i(x)$ has a dimension of ${\sf L}^{-c}$, ${\sf
  L}$ being the dimension of length, then the prefactor $x^{-c}$ takes
care of the dimension of the function, with $F_j$ taking care of the
additional dependence on $x$ and on $a$.  Since $F_j(z)$ has to be
dimensionless, its argument can only be $x/a$.  This leads to the form
of Eq. (\ref{eq:30}).  The exponent $-c$ that comes from dimensional
analysis is called the {\it engineering dimension} of $f_j$ 
(see Eq.~(\ref{eq:18})).

If we scale all lengths by a factor $b$, $x\to bx, a\to ba$, then
(keeping the $a$-dependence explicitly in the arguments)
\begin{equation}
  \label{eq:31}
f_{2,4}(bx,ba)=b^{-c}\; f_{2,4}(x,a),\;\; {\rm and\;}
f_3(x,a)=f_3(bx,ba),
\end{equation}
where the power of $b$ in the prefactor just reflects the power one
expects from dimensional analysis, its {\it engineering dimension}.
By choosing $b=1/x$, we recover the forms in Eq. (\ref{eq:30}). 

\subsection{Anomalous dimension}\index{dimension!anomalous}
\label{sec:anomalous-dimension}

In situations where $a$ refers to a small scale of the problem while
$x$ is a large length, e.g. $a$ may be the microscopic range of
interaction or lattice spacing while $x$ may be a macroscopic
distance, then $x/a\to\infty$ can be achieved by making $x\gg a$, or
even by taking $a\to 0$.  In this situation, naively $a$ may be set to
$0$, with $\tilde{f}_i$ approaching a constant.  Here we see,
$f_{2,4}(x\gg a)\sim x^{-c}$, while $f_3(x)\approx 0$.

There could be situations where $f(x)=x^{-c} {\tilde{f}}(x/a)$ is the
correct form with the engineering dimension but
${\tilde{f}}(z)\stackrel{z\to\infty}{\to} (x/a)^{-\eta}$, then for
$x/a\gg 1$, $f(x)\sim a^{\eta} x^{-c-\eta}$.  Most importantly, even
in the limit $x\gg a$, $a$ cannot naively be set to $0$.  The problem
is often stated in a dramatic way by setting $a=1$ to write $f(x)\sim
x^{-c-\eta}$ creating an {\it  illusion of violation} of the standard
dimensional analysis.  Consequently, this   additional exponent
$\eta$ is called the {\it anomalous dimension} of $f$.  It is more
natural to call $c+\eta$  the scaling dimension.

There is actually no violation of dimensional analysis as can be seen
by scaling both $x$ and $a$   because 
\begin{equation}
  \label{eq:46}
f(bx,ba)=b^{-c} f(x,a).  
\end{equation}
If we just scale $x$, the large length scale, without scaling the
intrinsic lengths like $a$, we get
\begin{equation}
  \label{eq:29}
f(bx,a)=b^{-c-\eta}\; f(x,{a}),
\end{equation}
which, by choosing $b=1/x$, says, $f(x,a)={\rm const} \ x^{-c-\eta}$ 
Note that scaling just $a$ gives 
\begin{equation}
  \label{eq:47}
  f(x,ba)=b^{\eta} f(x,a).   
\end{equation}
Eq.  (\ref{eq:47}) gives $f(bx,ba)=b^{\eta} f(bx,a)$, which can be
combined with Eq. (\ref{eq:46}), to write $f(bx,a)$ in the form of
Eq.~(\ref{eq:29}).  This suggests that the scaling behaviour of $f$
for large $x$, i.e., how the function changes as the variable $x$ is
changed by a scale factor, can be determined by combining dimensional
analysis (engineering dimension) with the changes expected as the
short distance scale is changed.\footnote{In many practical
  situations, $a$ plays the role of short distance cut-off.}

\subsection{Renormalization group flow equations}
\label{sec:renorm-group-flow}\index{renormalization group}

One way to generate the anomalous scaling behaviour is to obtain the
renormalization group (RG) flow equations.  Let us treat $a$ as a
continuous variable and take $b=1+\delta l$ so that $ba=a + \delta a$,
with $\delta a=a\; \delta l$.  Then Eq. (\ref{eq:47}), by Taylor
expansion, can be written as
\begin{equation}
  \label{eq:48}
  a\;\frac{\partial f}{\partial a}= \eta\; f.
\end{equation}
If we define a dimensionless quantity ${\hat f}=a^c\; f$, then by direct
differentiation with respect $a$,   and using Eq. (\ref{eq:48}), we
obtain
\begin{equation}
  \label{eq:49}
  a\;\frac{\partial {\hat f}}{\partial a}= c {\hat f}+ \eta\; {\hat f}.
\end{equation}
For $\eta=0$, the above equation is the expected equation with the
engineering dimension $c$ (compare with Eq. (\ref{eq:18})).  The extra
$\eta$-dependent term gives the anomalous contribution.  Such
equations that describe the change in the function as a microscopic
cut-off like variable (here $a$)  is changed, are called renormalization group
flow equations.  In general, in an RG flow equation, $\eta$ would be
dependent on the parameters of the problem, and only in special
situations (called {fixed points}), $\eta$ becomes a constant.
Under those conditions, i.e. at the fixed points, a proper power law
is obtained.   Proper scale invariance is observed at these fixed points.

\subsubsection{Examples of flow equations}
\label{sec:exampl-flow-equat}

As an example.  let there be two variables $v_1,v_2$ with
engineering dimensions $c_1, c_2$ respectively.  With an arbitrary
length $L$, the dimensionless parameters are $u_1=v_1 L^{-c_1}$, and
$u_2=v_2 L^{-c_2}$.  The $L$ dependence can be written in a form
analogous to Eq. (\ref{eq:16})
\begin{equation}
  \label{eq:32}
  L\frac{\partial u_1}{\partial L} = - c_1 u_1, \quad {\rm and}\quad 
  L\frac{\partial u_2}{\partial L} = - c_2 u_2.
\end{equation}
Now if it so happens that, due to  interactions or nonlinearities, the
actual $L$ dependence takes a form
\begin{equation}
  \label{eq:33}
  L\frac{\partial u_1}{\partial L} = - c_1 u_1 + b_1 u_1^2 + {\rm O}(u_1^3), \quad {\rm and}\quad
  L\frac{\partial u_2}{\partial L} = - c_2 u_2 + b_{12} u_1 u_2+ ... ,
\end{equation}
then $u_1$ attains a scale independent value at the fixed point
$u_1=0$ and $u_1^*=c_1/b_1$.  These are fixed points because $\partial
u_1/\partial L=0$.

Around $u_1=0$, it is the engineering dimension that matters even for
$u_2$, but at the nontrivial fixed point $u_1^*=c_1/b_1$, it seems
that $u_2$ acquires a new dimension $-{\hat c}_2=-c_2+\eta$, where
$\eta=b_{12}u_1^*=b_{12}c_1/b_1$.  This $\eta$ is the anomalous
dimension of $u_2$.  The idea of renormalization group (RG) is to
obtain equations like Eq. (\ref{eq:33}) to study deviations from
trivial behaviours.  ``Trivial'' here, of course, means results
obtained by dimensional analysis.

\begin{figure}[htbp]
\begin{minipage}{0.35\textwidth}
\begin{center}
\includegraphics[width=\textwidth]{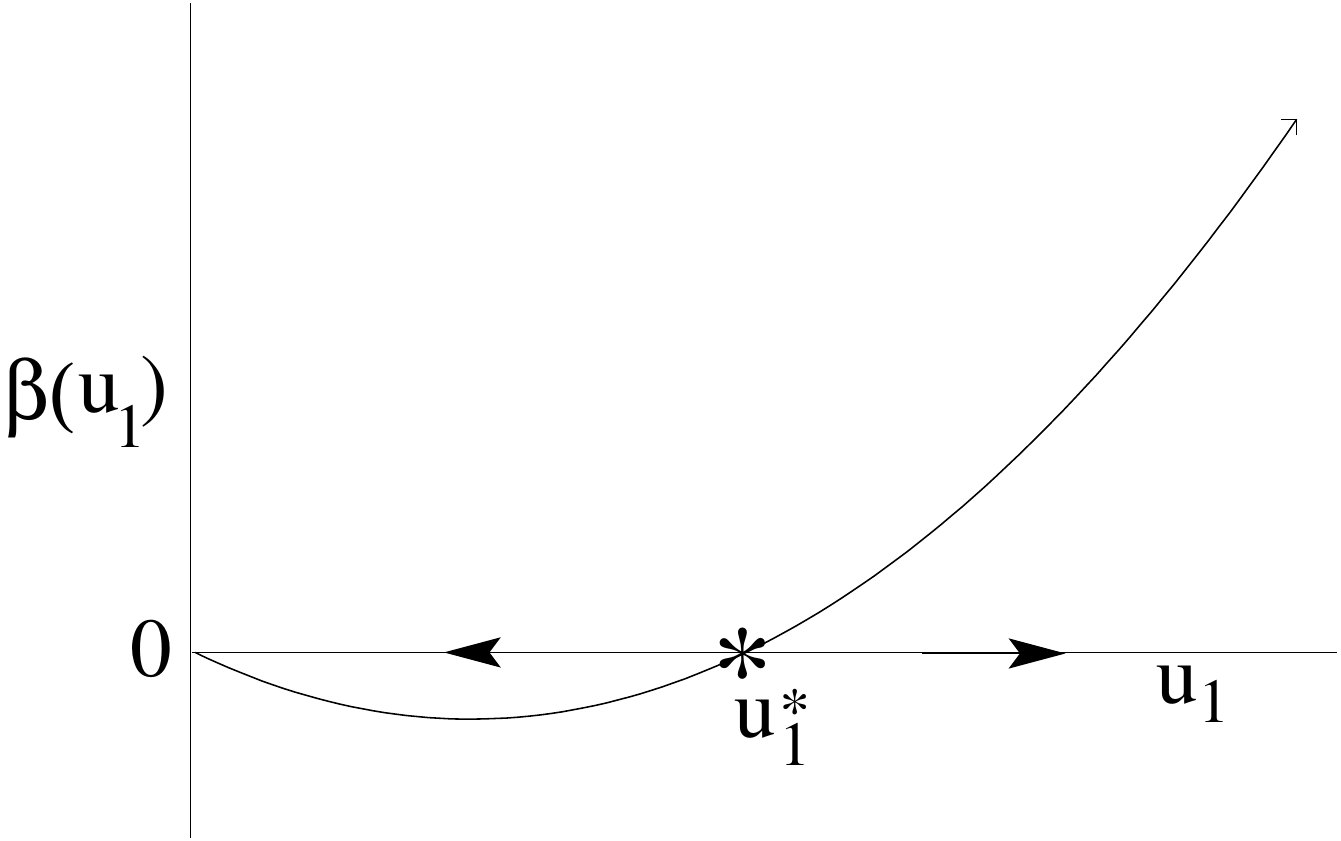}
\end{center}
\end{minipage}
\hfill\begin{minipage}{0.5\textwidth}
  \caption{RG flow of $u_1$.  The flow is described the
    $\beta$-function as given by Eq. (\ref{eq:33}).  The zero of the
    $\beta$-function, $u_1^*$ gives a critical point in this case as
    it is an unstable point.  The flows on the two sides of $u_1^*$
    are indicated by the arrows.  }\label{fig:betau}
\end{minipage}
\end{figure}

\subsubsection{Length scales from RG flow equations}
\label{sec:length-scales-from}

To elaborate on the renormalization group behaviour, we define the RG
flow equation as $L \partial u_1/\partial L=\beta(u_1)$.  For
concreteness, take $c_1,b_1>0$ with $u_1^*>0$.  The $\beta$-function
is shown in Fig. \ref{fig:betau}, with $u_1^*$ as a stable fixed
point, a zero of the $\beta$-function.  Any $u_1<u_1^*$ flows to zero,
while $u_1>u_1^*$ flows to infinity.  These  can be checked by a direct
integration of the flow equation for $u_1$.  Therefore, $u_1=u_1^*$ is
a critical point separating the two phases described by $u_1=0$ and
$u_1=\infty$.

The growth of $u_1$ away from the fixed point is also an  important
characterization of the function.    By linearizing around
$u_1=u_1^*$, with $\delta u=u_1-u_1^*$, 
\begin{equation}
  \label{eq:50}
L \frac{d\; \delta u}{d L} =
\left(\frac{d\beta}{du_1}\right)_{u_1=u_1^*} \delta u,   \quad {\rm
  or,}\quad \delta u= |u_{10}-u_1^*|\;\; (L/L_0)^{1/\nu},
\end{equation}
for the initial condition $u_1=u_{10}$ at $L=L_0$, and $\nu=\left[
  (d\beta/du_1)_{u_1=u_1^*}\right]^{-1}$.  We see that $\delta u$
reaches a preassigned value $\Delta$ at a length $\xi$, where
$\xi\sim |u_{10}-u_1^*|^{-\nu}$.  In other words, $\xi$ diverges as
$u_{10}$ approaches $u_1^*$.  The existence of a diverging length is
the hallmark of a critical point.  It is at this critical point, the
nontrivial fixed point in this example,  that  $u_2$
was found to acquire an anomalous dimension.

\begin{problem}
  Consider a particle in three dimensions in a central potential of
  the form (i) $V(r)=- A/(r+a)$, (ii) $V(r)=- \frac{e^{- \alpha
      r}}{r}$, each of which reduces to the attractive Coulomb
  potential for $a,\alpha\to 0$.  Discuss qualitatively the nature of
  the spectrum by comparing with the Hydrogen atom spectrum.
\end{problem}

\subsection{Example: localization by disorder - scaling of conductance}
\label{sec:exampl-local-disord}\index{conductance!scaling}\index{Gang of IV}

Let us consider the conductance $g(L)$ of a metallic sample in the
shape of a cube of side length $L$.  For small sizes, the conductance
of the sample is determined by the conductivity, $\sigma_0$.  A
macroscopic sample is obtained by successive rescaling $L$ to $2L$ and
so on.  Now, the conductance is due to the propagating electrons.  In
a pure metal (say a crystalline sample), the electrons are completely
delocalized as, e.g., described by the Bloch waves.  In contrast,
strong disorder, like impurities in the system, destroys the
translational symmetry, and can, instead, produce localized states for
the electrons.  If these are localized over a length $\xi$, then one
may observe some conductance for lengths $<\xi$ but not for $L\gg\xi$.
A question of importance is whether a macroscopic sample remains
metallic under disorder or there is a critical strength of disorder
beyond which a metal becomes an insulator.  A simpleminded RG approach
helps in answering this question.

\begin{figure}[h]
\begin{center}
\includegraphics[width=0.7\textwidth,clip]{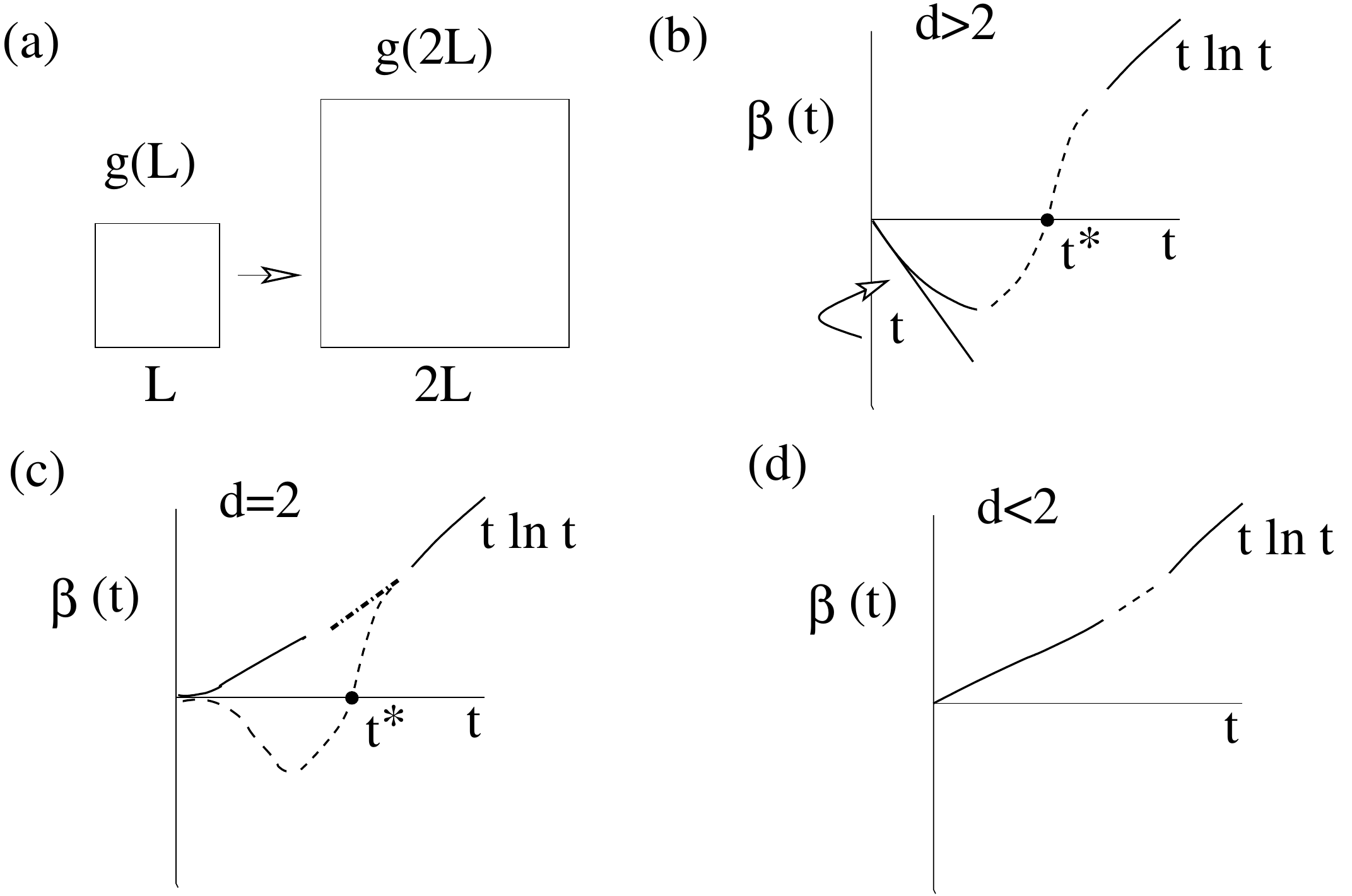}
\end{center}
\caption{(a) Rescaling of a hypercubic box of side length $L$ to $2L$.
  The conductance ($>0$) goes from $g(L)$ to $g(2L)$. (b) The RG beta
  function $\beta(t)$, for $d>2$, goes from $ \beta(t)\sim -t$ for
  small $t$ to positive values $\sim t \ln t,$ for large $t$.  The
  dotted line is a possible extrapolation which necessarily goes
  through a zero, $\beta(t^*)=0$. Here, $t^*$ is the RG fixed point.
  (c) For $d=2$, the $\beta$-function has zero slope at origin.
  Therefore, there are two possibilities, one with a zero and one
  without any. (d) For $d<2$, the slope at origin is positive, and
  there is unlikely to be any zero.  The dashed lines are possible
  extrapolations }\label{fig:giv}
\end{figure}

For a good conductor, we have $g(L)=\sigma_0 L^{d-1}/L\sim L^{d-2}$
for a $d$-dimensional hypercube, because the conductance is
proportional to the geometric factor $L^{d-1}/L$.  Defining $t=1/(2\pi
g)$, the scaling of $t$ can be expressed, in analogy with Eq.
(\ref{eq:18}) and Eq. (\ref{eq:32}), as \footnote{It is made
  dimensionless by the the universal constant $e^2/\hbar$, where is te
  electronic charge.}
\begin{equation}
  \label{eq:43}
  \frac{\partial \ln t(L)}{\partial  \ln L}=\epsilon\equiv (2-d), \quad {\rm or,}\quad L
  \frac{\partial t(L)}{\partial L} \equiv \beta(t) = \epsilon t,\quad ({\rm small}\quad t).
\end{equation}
This is the metallic regime.\footnote{This definition of
  the beta function follows the convention in statistical physics.
  The definition used in the original paper involves the log
  derivative.}

On the other hand, if the states are localized in the strong disorder
limit, i.e. all states are localized with a localization length $\xi$,
then the conductance can be expressed as $g(L)=g_0 \exp(-L/\xi)$.  In
this limit of $g\to 0$, and so we expect
\begin{equation}
  \label{eq:44}
L  \frac{\partial  g(L)}{\partial  L}=g\; \ln g,\quad (g\to0), \quad {\text{or,}}\quad L  \frac{\partial  t(L)}{\partial  L}=t\; \ln t,\quad (t\to \infty).
\end{equation}
By assuming that $\beta(t)$ depends only on $t$ (one parameter
scaling, as for $u_1$ in Eq. (\ref{eq:33}), we combine Eqs. \ref{eq:43} and \ref{eq:44} as
\begin{eqnarray}
  \label{eq:giv3}
  \beta(t)=\left \{ \begin{array}{ll}
                   \epsilon\; t, & {\text{for}}\; t\to 0,\\
                   t \ln t,       & {\text{for}} \; t\to \infty .
                   \end{array} \right.
\end{eqnarray}
We see that $\beta(t)$ for $d=3$ goes from a negative value for small
$t (>0)$ to a positive value, as shown schematically in Fig.
\ref{fig:giv}.  This means that there is a fixed point,
$\beta(t^*)=0$, where the parameter $t$ does not change with scale
(``scale invariant'').  It is straightforward to see that for any
initial value $t_0<t^*$, the flow equation on integration to large $L$
takes $t\to0$. Therefore on large scales the system behaves like a
metal.  In contrast for $t_0>t^*$ the flow goes to $t\to\infty$, an
insulator. The difference in behaviour on the two sides of $t^*$ is
ensured by $\beta'(t^*)\equiv d\beta(t)/dt|_{t^*}>0$.  The unstable
fixed point therefore represents a metal-insulator
transition.\index{conductance!flow equation}

The power of the RG flow equation can be seen in several ways.  If
there is a fixed point with $\beta'(t^*)>0$,i.e. an unstable fixed
point, it represents a transition point.  Since no such fixed point
exists for $d< 2$, the flow always goes to the insulator region.  In
other words, all states will be localized for $d<2$.  The linearized
form around the fixed point is
\begin{equation}
  \label{eq:giv1}
  L \frac{d\Delta t}{dL}\approx\beta'(t^*)\; \Delta t,\quad  \Delta t\equiv t-t^*.
\end{equation}
For small $\Delta_0\equiv |t_0-t^*|$, we may now find the length scale
$\xi$ at which $\Delta t$ reaches a predetermined value
$\tilde{\Delta}$.  On integration, Eq. \ref{eq:giv1} gives (omitting
the subscript of $t_0$)
\begin{equation}
  \label{eq:giv2}
  \xi\sim |t-t^*|^{-\nu},\quad{\text{where}}\quad
  \nu=\frac{1}{\beta'(t^*)}\sim \frac{1}{|\epsilon|}. 
\end{equation}
In fact, Eq.\ref{eq:giv2} is a very
general prediction from RG for any critical point.

\begin{problem}
  The disorder problem involves a Hamiltonian with random matrix
  elements.  Depending on the symmetry, the disorder problem can be
  classified in several groups with distinct $\beta$-function.  For
  each of the following $\beta$-functions, determine the fixed point
  and $\nu$, and discuss the behaviour in two dimensions (for
  $\epsilon=0$).
  \begin{enumerate}
  \item For a class, called the orthogonal symmetry class, $ \beta(t)=
    \epsilon t +2 t^2+...$.  Show that $t^*\approx |\epsilon|/2$ and $
    \nu=1/|\epsilon|, \epsilon<0$.  In two dimensions, the flow is
    towards the insulator side, i.e. toa state where all states are
    localized.

  \item For a class, called the unitary symmetry class, $\beta(t)=
    \epsilon t +2 t^3+...$.  Show that $t^*\approx
    \sqrt{|\epsilon|/2}, \nu=1/(2|\epsilon|), \epsilon<0$.  The two
    dimensional behaviour is same as the previous one, i.e., all
    states are localized.
  \item There is a class called the symplectic class for which
    $\beta(t)= \epsilon t - t^2+...$.  Show that at $d=2$, the
    behaviour is different from the above two, because now the flow is
    towards the metal side.
  \end{enumerate}
  For more details see F.  Evers and A. D. Mirlin, Rev.Mod.Phys. {\bf
    80}, 1355 (2008).
\end{problem}

\section{Multifractality}\index{multifractality}
\label{sec:multifractality}

For completeness we mention another idea, namely multifractality.

In the quantum mechanics context, the difference between the
wavefunctions of a bound state and an unbound state can be expressed
in terms of the finite size behaviour.  For a bound particle the size
$L$ of the box is not important so long it is much larger than the
width of the wavefunction.  For an unbound state $\psi\sim L^{-d/2}$.
These can be combined to write the behaviour of the finite size effect
of the moments of the wavefunction as
\begin{eqnarray}
  \label{eq:12}
  \int d^d r\; |\psi(r)|^{2q} \sim \left \{ \begin{array}{ll}
                                                     L^0,&{\rm (bound),}\\
                                                     L^{-d(q-1)},&{\rm (unbound),}\\
                                                     L^{-\tau_q},&
                                                       {\rm (critical),}
                                                       \end{array}\right. 
\end{eqnarray}
where we introduced a new class called ``critical'' wave function for
which $\tau_q$ is not linear in $q$ (note that the normalization condition
requires $\tau_1=0$).  For the bound state of width $\xi$, the
insensitivity of the boundary, for $L\gg\xi$, is expressed by the
power law $L^0$.  For extended states, the moments are completely
determined by the size of the box, with the exponent $d(q-1)$
following from dimensional analysis.  In contrast, for the critical
case, we find that for every moment a new length scale $\tau_p$ is
required so that a critical wave function requires an infinite number
of length scales to describe it.  It is generally written as
$\tau_q=d(q-1)+\Delta_q$, where $\Delta_q$ is the anomalous dimension.
In other words, moments of $\psi$ explore different aspects of how the
wave function is spread out in space.

For a large system, we may also revert to the box counting method of
Sec. \ref{sec:hausd-dimens-fract} as follows.  For a wave function
$\psi({\bf x})$, the probability $p(x)=|\psi(x)|^2$.  So in the box
method, instead of counting elements, we put an weight as
$$P_q= \sum_i \left(\int_{i\mathrm{th\  box}}
  d^d x\; |\psi({\bf x})|^{2}\right)^q = \sum_i \left(\int d^dx
  p(x)\right)^q,$$ 
where the summation is over all the boxes of size $b$ covering the
sample.\footnote{In cases involving disorder, as in the localization
  problem, a sample averaged quantity $[P_q]_{\mathrm{dis}}$ is to be
  calculated, where $[...]_{\mathrm{dis}}$ denotes an averaging over
  samples of disorder.  For convenience, we drop the averaging
  symbol.}  There are $n=(L/b)^d$ number of boxes.  Necessarily,
$P_1=1, P_0=n$.  In analogy with Eq. (\ref{eq:12}), we define
\begin{equation}
  \label{eq:56}
  P_q\sim \left(\frac{L}{b}\right )^{-\tau_q}.
\end{equation}
In such a situation, the quantity of interest is the fractal dimension
of the set of points where $p=|\psi|^2 \sim L^{-\alpha}$.  Let the
fractal dimension be given by $f(\alpha)$, i.e., the measure of the
set of point with $|\psi|^2\sim L^{-\alpha}$ is $L^{f(\alpha)}$.  This
fractal dimension depends continuously on $\alpha$, justifying the
name of ``multifractal''.  We just state here that the multifractal
spectrum $f(\alpha)$ is related to $\tau_q$ by a Legendre
transformation\footnote{See Ref \cite{janssen} for details. } as
$$f(\alpha)=q\alpha - \tau_q, \quad {\mathrm{  where}}\quad 
\alpha=\partial \tau_q/\partial q.$$  The wavefunction at the localization transition mentioned in
the Sec.  \ref{sec:exampl-local-disord} is an example of a
multifractal wave function.

A tractable example of multifractal-like behaviour would be a power
law decay of a wave function $\psi(r)$ which for large $r$ behaves as
$\sim r^{-u}$, where $ 1/2<u<1$ In this situation, since $\int dr
|\psi|^2$ is convergent, the wave function is normalizable, and
therefore represents a bound state.  The unusual nature of the bound
state can be seen from the behaviour of the $q$th moment given by
$\int dr \;r^q\; |\psi|^2$ which is divergent for $q> 2u-1$.  Note that $q$
is not necessarily an integer.  A bound state with energy $E<0$ gives
a length scale, $\xi\sim1/\sqrt{|E|}$ which one may associate with the
scale beyond which the wave function decays exponentially as
$\exp(-r/\xi)$.  In a sense this gives the width of the wave function
and for most cases, like the square well potential or short range
potentials, this scale is enough.  However the situation we are
considering corresponds to the case where for $\xi\to\infty$, the wave
function goes over to the power law decay so that for $E$ very close
to zero, there will be an intermediate range where the power law form
is visible;   for example a form like $\psi(r)\sim
r^{-u}\; {\exp(-r/\xi)}$.  A length scale to characterize the
wavefunction in this intermediate range can be obtained from the
moments as $\int dr \;r^q\; |\psi|^2\sim\int^{\xi} dr\; r^{q-2u}\sim
\xi^{q-2u+1}$, for $\xi\to\infty$.  In this limit the normalization
constant is a number independent of $\xi$.  We therefore find
$l_q\sim\xi^{\tau_q}$ where $\tau_q=1 - (2u-1)/q$, with an anomalous
exponent $D_q=-(2u-1)/q$.  This is not a carefully crafted example
but occurs at the unbinding transition of a quantum particle in a
potential $V(r)+a/r^2$ in three dimensions where $V(r)$ is a short
range attractive potential.  By tuning the short range potential, a
zero energy bound state can be formed whose wave function decays in the
power law fashion just mentioned.

\section{Conclusion}

In this chapter, we explored various definitions of dimensions, like
the topological, the Hausdorff, the box or Minkowski, and the spectral
dimensions by embedding the set in Euclidean space.  When the Hausdorff 
and the box dimensions are the same, it is
called the fractal dimension, and the set is called a fractal if this fractal
dimension  is different from the topological dimension.  For Euclidean spaces, 
all these definitions give the same number, which, by construction, is a 
positive integer.
However we have explicitly constructed various subsets of the
Euclidean space whose dimensions are not integers.  Further
generalizations are made to study power law
behaviour of various physical quantities in terms of renormalization
group flow equations.   The flow equations show the emergence of
anomalous dimensions  at certain special fixed points as opposed to
engineering dimension determined by dimensional analysis.
We also discussed how various physical
properties are determined by various dimensions of the space.

\section*{Appendix A: Entropy and fractal dimension}\label{sec:entr-fract-dimens}
\addcontentsline{toc}{section}{Appendix A: Entropy and fractal dimension}
\index{entropy}

We here establish the relation between entropy per spin of a chain of
spins and the fractal dimension of the configuration space when mapped
on to real numbers.  The discussion is  for a semi-infinite chain
of spins which can be labeled by integrers, $1,2,...$. 
 
Let us consider a chain  of spins, each taking two values, $0$,
and $1$.  If all configurations are equally likely to occur, then the
number of configurations for $N$ spins is $C_N=2^N$.   The 
entropy per spin is therefore $s= k_B (\ln C_N) /N=  k_B \ln 2$.
In general, the entropy per spin can be written as a derivative
\begin{equation}
  \label{eq:64}
s=k_B \;\;\frac{\ln C_{N+1}- \ln
C_N}{(N+1)-N}=k_B \;\; \left. \frac{\partial \ln C_N}{\partial N}\right|_{N\to\infty},  
\end{equation}
where the last term is in the continuum limit.

Now we convert the strings of $0$ and $1$, $(s_n=0,1|n=1,...,N)$, to a
real number by using base $m=3$ as 
\begin{equation}
  \label{eq:62}
x=\sum_{n=1}^N  \frac{2s_n}{m^n}.    
\end{equation}
As
discussed in  Sec.  \ref{sec:conf-space-ising}, for
$N\to\infty$, $x$ forms the Cantor set.  
We chose $3$ because of our familiarity with the Cantor set, but $m=3$
is nothing special.

The fractal dimension of the subset of real numbers generated by all
the spin configurations is given by the box dimension Eq.
(\ref{eq:57}).  As we go from $N$ to $N+1$ in the number of spins, we
add a higher order term $2s_{N+1}/m^{N+1}$ in $x$.  This is equivalent
to changing the scale of the box size from $m^{-N}$ to $m^{-(N+1)}$,
the latter being the finer scale.  In other words, the ``box''size has
been changed by a scale factor $1/m$.  The denominator of Eq.
(\ref{eq:57}), with $b=1/m$, becomes $ \ln m^{N+1}-\ln
m^{N}=[(N+1)-N]\ln m $.  The fractal dimension of the set is
therefore
$$\fbox{$\displaystyle{d_f=\frac{1}{\ln m}\;\; \frac{\ln C_{N+1}- \ln
      C_N}{(N+1)-N}
=\frac{1}{\ln m}\;\;\frac{s}{k_B},}$}$$
where we used Eq. (\ref{eq:62}).
This establishes the connection between the entropy per spin and the
fractal dimension of the set of real numbers equivalent to the configuration
space.

For the Ising case, with $m=3$, we see $s\propto \ln 2/\ln 3=\log_3 2$.

One may wonder, why we chose base $3$ ($m=3$).  In general, for any
choice of $m\geq 2$, the real numbers 
\begin{equation}
  \label{eq:65}
x=\sum_{n=1}^N\frac{(m-1)s_n}{m^n},  
\end{equation}
form a subset of $ [0,1]$.  As the above
result shows, we could have chosen any $m$ to get $s\propto \ln 2/\ln
m$.  The extra factor $\ln m$ could easily be absorbed in $k_B$ which
is equivalent to changing the base of the logarithm.  Our choice of
$m=3$ is motivated by the fact that it is the smallest integer for
uniqueness of the mapping, and our familiarity with the Cantor set.

Let us clarify the problem of mapping by taking the simplest situation
of base $2$ ($m=2$).   
For any string we define $x=\sum_{i=1}^{\infty} s_n/2^n$, so that
$x\in[0,1]$.  Note that  we get the whole interval.  
However, two strings $s_1s_2...s_n011111....$ gives the
same value of $x$ as $s_1s_2...s_n1000000....$, like in traditional
decimal system where $1.0=0.9999999...$.  Therefore with base 2, we
get a mapping from the binary strings to the real numbers $\in[0,1]$,
but it is not unique; this is a many to one mapping.
In contrast, with base $m\geq 3$, we get a unique one-to-one, onto and
invertible mapping via Eq. (\ref{eq:65}).
Nevertheless, as Prob 4.3 shows, the entropy can be related to the
dimension of the space of points generated out of the strings with any
base $m$.  This includes $m=2$ with $\log_2 2=1$, the dimension of the
interval $[0,1]$!

In fields, like computing and telecommunication, $\log_2 2=1$
is used as a unit (called Shannon) of information content  of one  bit (same as entropy).  
\footnote{See Ref.\cite{smbent} for a
discussion on definitions of entropy.}

The connection between entropy and fractal dimension  for more general
situations of spin chains are given
as problems (Prob 4.1,4.2,4.3).   Whether this connection can be
extended to more general systems or general lattices  remain to be seen.  

\section*{Appendix B: Complex dimension: continuous and discrete Scaling}
\label{sec:complex-dimension}
\addcontentsline{toc}{section}{Appendix B: Complex dimension:
  continuous and discrete Scaling}

This appendix is technical in nature and may be skipped without loss of
continuity.\index{dimension!complex}

The self similarity discussed so far is of geometric nature.   This may
be true for any property of a system.  If a function $f(x)$ satisfies a relation
$f(x)=b^{\mu} f(bx)$ then $f(x)$ is said to be scale invariant as $b$
may be viewed as a scale factor for the variable.  If $b$ is
arbitrary, then we may choose $bx=1$ to obtain $f(x)\sim x^{-\mu}$, a
power law dependence on $x$.  Thus power laws are synonymous to scale
invariance - something  one  sees near critical points.  Since $b$ is
arbitrary, such a scale invariance is called a continuous scale
invariance.  

The scale invariance we saw for the geometric fractals are not
continuous but discrete.  Instead of choosing powers of $3$, suppose
we choose some other $b, 3^{n}<1/b<3^{n+1}$ as the scale factor for
the Cantor set.  For such a scale factor $x=1/b$, the number of pieces
would remain the same, changing only when x matches with the correct
scaling factor.  It is then possible to write
\begin{equation}
  \label{eq:19}
  C_x=C x^{d_{\rm f}} F\left(\frac{\ln x}{\ln 3}\right),
\end{equation}
where $F(z)$ is a periodic function of periodicity $1$.
A Fourier expansion gives  
\begin{equation}
  \label{eq:21}
  F(z)=\sum_n ( a_n e^{i2\pi n z}+a_n^* e^{-i2\pi n z}),
\end{equation}
where $*$ denotes complex conjugation.  On substitution in
Eq. (\ref{eq:19}), one gets a simpler power law form with
\begin{equation}
  \label{eq:22}
  d_{\rm f}^{\prime}= d_{\rm f}+ i \frac{2\pi n}{\ln 3}, \quad n\in \mathbb{Z},
\end{equation}
a tower of complex dimensions. Restricting to the first mode, the
oscillatory behaviour is of the form 
\begin{equation}
  \label{eq:23}
C_x\sim x^{d_{\rm f}} \cos\left(2\pi \frac{\ln x}{\ln 3}\right).   
\end{equation}
Such an oscillatory behaviour for an arbitrary scale factor is a
distinct signature of a discrete scale invariance.  Whether these will
have any important perceptible effect ultimately depends on the
amplitudes $a_n$.  In many situations, $|a_n|$ turns out to be
extremely small compared to the nonoscillatory terms.

A notable example of a continuous scale invariance breaking into a
discrete one is the Efimov effect in three body quantum mechanics or
its classical analog in three stranded DNA.

\subsection*{B.1 Cantor string}
\label{sec:cantor-string}\index{Cantor string}
\addcontentsline{toc}{subsection}{B.1 Cantor string}

Consider the complement of the Cantor set in the closed interval
$[0,1]$.  This is the set of disjoint lengths (open intervals) which
add up to a length $1$, but still it is the whole line segment minus
the set of points belonging to the Cantor set.  A bounded open set of
$\mathbb{R}$ is a fractal string, and the particular one we are
discussing is the Cantor string.
A poetic name is a one-dimensional drum with fractal boundary.  A
relevant question  is "Does one hear the shape of a drum?''

The fractal sting is described by the set of lengths $l_j$,
$j=1,\infty$.  For the Cantor string, these are
$\frac{1}{3},\frac{1}{9},\frac{1}{9},...$, keeping track of the
multiplicities (i.e., degeneracies), length $l_j$ occurring $m_j$
times.    Let us define a zeta function 
\begin{equation}
  \label{eq:41}
  \zeta(s)=\sum_{j=1}^{\infty}\; m_j l_j^s,
\end{equation}
where $s$ is a complex number so that the series is convergent. For
$s=1$, $\zeta(1)=1$, the length of the string.  Since
$l_j<1$, the series is definitely convergent for large positive real $s$.  The
minimum real value of $s$ for which it is convergent happens to be the
fractal dimension of the boundary.  On analytic continuation, one
may define $\zeta(s)$ over the complex $s$-plane with singularities
which are the complex dimensions of the boundary set.  With an abuse
of definition, the fractal dimension of the boundary is also called the
dimension of the string.

For the Cantor string,  $l_j=\frac{1}{3^j}$ with degeneracy
$m_j=2^{j-1}$, so that
\begin{equation}
  \label{eq:42}
  \zeta(s)=\sum_{j=1}\frac{1}{3^s}\;
  \left(\frac{2}{3^{s}}\right)^{j-1}=\frac{3^{-s}}{1-(2\times 3^{-s})},
\end{equation}
which has poles at $3^s=2$, with $s$  given by Eq. (\ref{eq:22}).

Why should such a zeta function be useful?  This becomes clear if we
look upon Eq. (\ref{eq:41}) as a transformation for the weights $w_j$.
Since $l_j<1$ and $l_j\to 0$ for $j\to\infty$,  one may, in a very nonrigorous
way,  write the function as an integral
\begin{equation}
  \label{eq:45}
  \zeta(s)=\int_0^{\infty} \ x^{-s} w(x) dx,
\end{equation}
identifying the zeta function as the Mellin transformation of the
weight function.  A quantity of interest is the number of intervals of
size less than ${\it x}$, $N(x)\sim \int_0^x w(x) dx$ for which the
zeta functions are useful.

\section*{Appendix C: Spectral dimension for the Sierpinski Gasket}
\label{sec:spectr-dimens-sierp}
\addcontentsline{toc}{section}{Appendix C: Spectral dimension for the Sierpinski Gasket}

A scalar phonon problem  on the Sierpinski gasket  involves springs along the  bonds
of with equal masses at the sites but the restoring forces are added
added disregarding the vectorial nature of the forces.   This is
equivalent to a resistor problem of finding the equivalent resistance
between two sites if all the bonds are occupied by 1 Ohm resistors.
By Kirchoff's law, all the voltages are linearly
added.\index{Sierpinski Gasket!spectral dimension}

\begin{figure}[htbp]
\begin{minipage}[c]{0.47\textwidth}
\includegraphics[scale=0.75,clip]{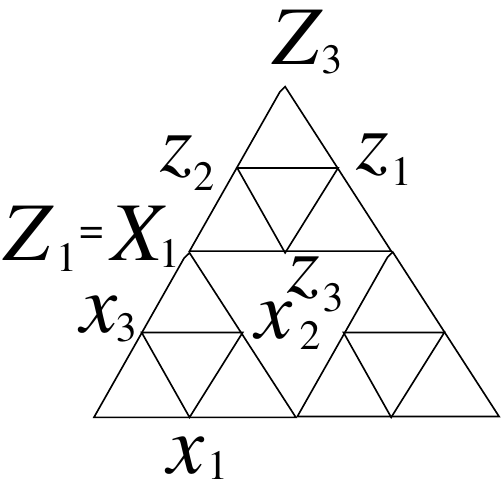}
\end{minipage}\hfill
  \begin{minipage}[c]{0.5\textwidth}
    \caption{Notations for displacements or voltages in the scalar
      phonon problem.  The capital letters will survive on elimination
      of the small letter variables from the equations.
    }\label{fig:sierspring}
\end{minipage}
\end{figure}

With the notation $\lambda=m \omega^2/K$ where $m$ is the mass at the
sites and the spring constant on the bonds, the equation of motion for
say $x_1$ is
\begin{equation}
  \label{eq:34}
  (4-\lambda) x_1 - x_2-x_3= X_2+X_3,
\end{equation}
while the equations for  $x_2,x_3$ are obtained by appropriate
permutations of the variables.  Similarly for $z_i$'s.  For the
capital variables,
\begin{equation}
  \label{eq:35}
  (4-\lambda) X_1= x_2 +x_3+ z_2+z_3,
\end{equation}
and so on.   It is now straightforward to eliminate $x_i$'s and
$z_i$'s to obtain an equation involving only the capital variables, as
\begin{equation}
  \label{eq:36}
  (4=\lambda')X_1=X_2+X_3+Z_2+Z_3, \quad \lambda'=\lambda(5-\lambda).
\end{equation}
so that for small frequencies, $\lambda'=5\lambda$.

Because of the change in the number of degrees of freedom under this
scale change by $b=2$, the density of states should change as
\begin{equation}
  \label{eq:37}
  \rho_{L/b}(\omega)=b^{-d_{\rm f}} \rho_L(\omega),
\end{equation}
while the frequency itself may scale as
\begin{equation}
  \label{eq:38}
  \omega(L)=b^{-\kappa} \omega(L/b).   
\end{equation}
Since the number of states remain
  invariant, i.e., $\rho_{L/b}(\omega') d\omega'= \rho_L(\omega)
  d\omega$, we have (using Eqs. (\ref{eq:37},\ref{eq:38}))
  \begin{equation}
    \label{eq:39}
    \rho_{L/b}(\omega)=b^{-\kappa} \rho_L(\omega b^{-\kappa})=b^{-d_{\rm f}}\rho_L(\omega).
  \end{equation}
If we choose $\omega b^{-\kappa}=1$, then $\rho_L(\omega)=\omega^{d_{\rm f}/\kappa
  -1}$.  The spectral dimension is therefore $d_s=d_{\rm f}/\kappa.$
 
From $\lambda'\approx 5 \lambda$,  we have $z^{2\kappa}=5$, or, $\kappa=\frac{\ln
  5}{2\ln 2}\neq 1$.  Combining all, we find the spectral dimension of
the Sierpinski Gasket to be 
\begin{equation}
  \label{eq:40}
  d_s=\left(\frac{\ln 3}{\ln 2}\right) \Big/ \left(\frac{\ln 5}{2\ln 2}\right)= 2 \;  \frac{\ln 3}{\ln
    5},
\end{equation}
using  the fractal  dimension, $d_{\rm f}=\ln 3/\ln 2$.

\newcommand{\finaladr}{%
\vspace{8pt} 
\hrule
\vspace{6pt} 
\noindent{\textsc{{Somendra M. Bhattacharjee, Institute of Physics, Bhubaneswar 751 005 
India, and Homi Bhabha National Institute, Training School Complex, Anushakti Nagar, Mumbai 400085, India\\
  email:}}somen\string@iopb.res.in
} 
\vspace{6pt} 
\hrule
}%

\end{document}